\documentclass[12pt,english]{article}
\usepackage[T1]{fontenc}
\usepackage[latin9]{inputenc}
\usepackage{geometry}
\usepackage{color}
\usepackage{babel}
\usepackage{refstyle}
\usepackage{url}
\usepackage{amsmath}
\usepackage{amsthm}
\usepackage{amssymb}
\usepackage{graphicx}
\usepackage{setspace}
\usepackage[authoryear]{natbib}
\usepackage[unicode=true,
 bookmarks=true,bookmarksnumbered=true,bookmarksopen=true,bookmarksopenlevel=1,
 breaklinks=false,pdfborder={0 0 0},pdfborderstyle={},backref=false,colorlinks=true]
 {hyperref}
\hypersetup{
 citecolor=blue, linkcolor=blue}

\makeatletter

\AtBeginDocument{\providecommand\axmref[1]{\ref{axm:#1}}}
\AtBeginDocument{\providecommand\subsecref[1]{\ref{subsec:#1}}}
\AtBeginDocument{\providecommand\figref[1]{\ref{fig:#1}}}
\AtBeginDocument{\providecommand\oappxref[1]{\ref{oappx:#1}}}
\AtBeginDocument{\providecommand\exaref[1]{\ref{exa:#1}}}
\AtBeginDocument{\providecommand\secref[1]{\ref{sec:#1}}}
\AtBeginDocument{\providecommand\appxref[1]{\ref{appx:#1}}}
\AtBeginDocument{\providecommand\propref[1]{\ref{prop:#1}}}
\AtBeginDocument{\providecommand\corref[1]{\ref{cor:#1}}}
\AtBeginDocument{\providecommand\thmref[1]{\ref{thm:#1}}}
\AtBeginDocument{\providecommand\defref[1]{\ref{def:#1}}}
\AtBeginDocument{\providecommand\condref[1]{\ref{cond:#1}}}
\AtBeginDocument{\providecommand\eqref[1]{\ref{eq:#1}}}
\RS@ifundefined{subsecref}
  {\newref{subsec}{name = \RSsectxt}}
  {}
\RS@ifundefined{thmref}
  {\def\RSthmtxt{theorem~}\newref{thm}{name = \RSthmtxt}}
  {}
\RS@ifundefined{lemref}
  {\def\RSlemtxt{lemma~}\newref{lem}{name = \RSlemtxt}}
  {}

\numberwithin{equation}{section}
\numberwithin{figure}{section}
\theoremstyle{plain}
\newtheorem{ax}{\protect\axiomname}
\theoremstyle{definition}
\newtheorem{defn}{\protect\definitionname}
\theoremstyle{plain}
\newtheorem{thm}{\protect\theoremname}
\theoremstyle{plain}
\newtheorem{prop}{\protect\propositionname}
\theoremstyle{definition}
 \newtheorem{example}{\protect\examplename}
\theoremstyle{definition}
\newtheorem{condition}{\protect\conditionname}[section]
\theoremstyle{remark}
\newtheorem*{claim*}{\protect\claimname}
\theoremstyle{plain}
\newtheorem{lem}{\protect\lemmaname}
\theoremstyle{plain}
\newtheorem{cor}{\protect\corollaryname}

\usepackage{amssymb}
\usepackage{refstyle}
\usepackage{accents}
\usepackage{charter}
\usepackage{paralist}

\def\sloppy{%
  \tolerance 1000%
  \emergencystretch 1em%
  \hfuzz .5\p@
  \vfuzz\hfuzz}
\newref{sec}{name=Section~}
\newref{subsec}{name=Subsection~}
\newref{axm}{name=Axiom~}
\newref{lem}{name=Lemma~}
\newref{def}{name=Definition~}
\newref{prop}{name=Proposition~}
\newref{thm}{name=Theorem~}
\newref{remark}{name=Remark~}
\newref{cor}{name=Corollary~}
\newref{fig}{name=Figure~}
\newref{eq}{name=Equation~}
\newref{appx}{name=Appendix~}
\newref{oappx}{name=Online Appendix~}
\newref{fn}{name=Footnote~}
\newref{fact}{name=Fact~}
\newref{exa}{name=Example~}
\newref{cond}{name=Condition~}
\newref{claim}{name=Claim~}

\makeatother

\providecommand{\axiomname}{Axiom}
\providecommand{\claimname}{Claim}
\providecommand{\conditionname}{Condition}
\providecommand{\corollaryname}{Corollary}
\providecommand{\definitionname}{Definition}
\providecommand{\examplename}{Example}
\providecommand{\lemmaname}{Lemma}
\providecommand{\propositionname}{Proposition}
\providecommand{\theoremname}{Theorem}

\begin{document}
\title{\textbf{Choice and Attention across Time}}
\author{Xi Zhi Lim\thanks{Shanghai Jiao Tong University (email: xzlim@sjtu.edu.cn). I am grateful
to co-advisors Mark Dean and Pietro Ortoleva for guidance and to committee
members Hassan Afrouzi, Navin Kartik, and Qingmin Liu for feedback.
I thank Kyle Chauvin, Xiaoyu Cheng, Paul Cheung, Soo Hong Chew, David
Dillenberger, Evan Friedman, Shaowei Ke, Matthew Kovach, Jian Li,
Senran Lin, Shuo Liu, Yusufcan Masatlioglu, Bin Miao, Daisuke Nakajima,
Norio Takeoka, Rui Tang, Fan Wang, Chen Zhao, seminar participants
at AMES 2022, CMES 2021, Columbia, MEET 2022, NYU Shanghai, Peking
University, RUD 2023, SHUFE, SWUFE (CCBEF), and VEAEBES for feedback.
This work appears in the 3rd chapter of my PhD dissertation at Columbia
University (2020).}}
\date{\vskip -.8emThis version: 2024/03/02\\
Most recent and public: \url{http://s.xzlim.com/sequence}}
\maketitle
\begin{abstract}
I study how past and future choices are linked in the framework of
attention. Attention cannot be observed but past choices are necessarily
considered in future decisions. This link connects two types of rationality
violations, counterfactual and realized, where the former results
from inattention and the latter fully pins down preferences. Results
show that the necessary traces of limited attention lie within choice
sequences because they enable and compel a decision maker to correct
their \textquotedblleft mistakes\textquotedblright . The framework
accommodates different attention structures and extends to framing,
introducing choice sequences as an important channel to formulate,
identify, and scrutinize limited attention.\\
\\
\textbf{Keywords}: Choice sequences, limited attention, limited consideration,
framing effects\\
\textbf{JEL}: D01, D11 
\end{abstract}

\newpage{}

\section{Introduction}

The generalization of a standard theory to explain ``non-standard''
behavior benefits from richer data; intuitively, data can compensate
for the added flexibility. Observing reference points, endowments,
or status quo reconciles behaviors that contradict a single, consistent
preference.\footnote{For example, \citet{kahneman1979prospect,munro2003theory,sugden2003reference,masatlioglu2005rational,ortoleva2010status,masatlioglu2013choice,masatlioglu2014canonical,dean2017limited,kovach2020twisting,ellis2022choice}.}
Observing menu preferences helps test the hypothesis of temptation
and enables studies of self-control and addiction.\footnote{For example, \citet{gul2001temptation,gul2007harmful,dekel2009temptation,noor2011temptation,dillenberger2012ashamed,ahn2019behavioural,freeman2021revealing}.}
Observing compound lottery and multi-dimensional risk brings new insights
to traditional risk preference anomalies like the Allais paradox.\footnote{For example, \citet{segal1990two,dillenberger2010preferences,lanzani2022correlation,chew2022source,halevy2022uncertainty,zhang2023procedural,ke2023multidimensional}.}
These datasets are appreciated because they are useful and innovative,
even if they change the ways data have to be collected. Can a dataset
that is useful but \emph{ubiquitous} also receive consideration? Stochastic
choice is one example,\footnote{For example, \citet{gul2006random,manzini2014stochastic,gul2014random,fudenberg2015stochastic,brady2016menu,aguiar2017random,echenique2018perception,cattaneo2020random,kovach2022behavioral,kovach2023reference,kibris2024random}.}
choice sequences could be another.

The growing literature on limited consideration has thus far\emph{
zoomed in} on a choice problem to study an attention mechanism, investigating
the possibility of a search process that considers a subset of alternatives
or a rule of thumb that eliminates alternatives from final decisions.
This paper \emph{zooms out} and studies how the evolution of choices
can hint at the role of attention even if no assumptions are imposed
on attention structures.

Limited consideration occurs when a decision maker (DM) fails to consider
every alternative in every choice set. It results in seemingly irrational
decisions even if the DM has a standard and consistent preference,
thereby capturing a straightforward form of bounded rationality that
has received substantial attention.\footnote{Earliest work traceable to \citet{wright1977}'s discussion of consideration
set, related applications in marketing and finance that include \citet{hauser1990evaluation,roberts1991development},
and related choice theories that include \citet{manzini2007sequentially,masatlioglu2012revealed,cherepanov2013rationalization}.} But its simplicity and intuitiveness is not without cost. Attention
being inherently hard to observe can leave us with multiplicity of
estimated preferences, which burdens our analysis of economic consequences
and welfare. Moreover, the model specification that $x$ receives
attention in choice problem $A$ but not in choice problem $B$ is
almost impossible to test with a within-subject design; once the DM
experiences $A$ and pays attention to $x$, would she immediately
forget $x$ when asked to choose from $B$?

This paper addresses these issues by linking past and future choices,
allowing an analyst to exploit the wealth of information contained
in the natural evolution of choices. The innovation lies in the primitive\textemdash a
dataset of choice sequences, which departs from the standard ``one-shot''
setting where the DM makes one real choice.

This is accompanied by an intuitive assumption: past choices should
be automatically considered when the DM makes future decisions. Then,
decisions from the same choice set that vary with experience hint
at a mechanism of evolving attention. Moreover, because choices must
become more informative, a latter choice that contradicts an earlier
choice reveals true preference. This contributes to necessary pleasantries:
either behavior is standard, or its ``non-standard'' manifestation
reveals unobserved parameters. It turns out that this simple assumption,
captured using an axiomatic foundation, allows us to test the hypothesis
of limited consideration using realized choices, confirm ``mistakes'',
and fully identify preferences.

To illustrate, suppose you are unaware of the vacation destination
Penang (an island in my country Malaysia) even though you can afford
it, so an analyst who observes your choice of Hawaii may falsely jump
to the conclusion that you prefer Hawaii over Penang by the theories
of revealed preferences. However, when you attend a conference in
Southeast Asia, you might consider and choose Penang for a drop-by
vacation. This incident makes Penang then and forever an option you
are aware of, and your future decisions of whether to return to Penang
will more informatively convey your true preference between Penang
and other destinations.

The underlying intuition applies broadly: A DM who uses the iPad may
or may not have considered a Surface Go, but a DM who converted to
an iPad from a Surface Go probably prefers iPad to Surface Go. A person
who reads physical books may actually prefer e-readers, but one who
left e-readers for physical books probably prefers physical books.
A colleague who has not begun to referee papers could be a remarkable
reviewer, but one who used to do so yet is no longer invited may not
be the best reviewer.

To learn from this intuition, consider a DM with an \emph{Attention
Across Time} (AAT) representation. The DM has a subjective attention
function that determines what she would normally consider from each
choice set $A$, denoted by $\Gamma\left(A\right)$. Moreover, the
DM's past experiences, $h$, identifies a collection of previously
chosen alternatives that will continue to be considered, denoted by
$\boldsymbol{c}\left(h\right)$. The DM's decision therefore solves
\[
\tilde{c}\left(h\right)\left(A\right)=\max_{x\in\Gamma\left(A\right)\cup\left(\boldsymbol{c}\left(h\right)\cap A\right)}\,u\left(x\right).
\]
Of course, past experiences $h$ are historical choice problems, so
elements of $\boldsymbol{c}\left(h\right)$ come from the same procedure,
just at an earlier time. This contrasts the ``standard'' DM who
has full attention $\Gamma\left(A\right)=A$, for whom the problem
reduces to the familiar
\[
\tilde{c}\left(h\right)\left(A\right)=\max_{x\in A}\,u\left(x\right).
\]

The first inquiry helps us understand the behavioral content of this
model; three key axioms that relate past and future choices underpin
AAT behavior. \emph{Weak Stability} (\axmref{1}) imposes restriction
on behavior over time: in contrast to full compliance with WARP, it
allows for one-time switches between every pair of alternatives. \emph{Past
Dependence} (\axmref{2}) limits the way past choices may affect future
choices; specifically, if a recent experience affects the choice from
the current choice set, then the new choice is limited to the recently
chosen alternative. The third and last axiom captures the behavioral
signature of attention. First, revealed preference\emph{ }is defined
when observed choices suggest that the DM is aware of $y$ when $x$
is chosen. Specifically, when $x$ is chosen over $y$ after $y$
was previously chosen, or when $x$ is chosen from a choice set where
$y$ is chosen ``by default''. \emph{Default Attention} (\axmref{3})
posits that if $x$ is revealed preferred to $y$, then $y$ will
never be chosen from a choice set where $x$ is chosen by default.
The axiom essentially says that the default choice from a choice set
should receive attention in that choice set no matter the history,
and therefore a subjectively inferior alternative should never be
chosen. All three axioms are trivially satisfied in the conventional
setting with only one period of choice.

Then, a series of straightforward observations forms the core of this
paper: exploring what economists can learn from the wealth of information
in choice sequences.

The first observations concern the identification of preferences.
Preferences are pinned down. To illustrate the intuition, suppose
a WARP violation occurs between the choice of $x$ from $\left\{ x,y\right\} $
and the choice of $y$ from $\left\{ x,y,z\right\} $. It turns out
that if we present the ``problematic'' choice sets in an alternating
order, then the DM is bound to consider both $x$ and $y$ in the
second problem, resulting in a choice that reveals their preference.\footnote{If genuine preference is $u\left(x\right)>u\left(y\right)$, then
the sequence of choice sets $\left(\left\{ x,y\right\} ,\left\{ x,y,z\right\} \right)$
produces choices $\left(x,x\right)$ and the sequence of choice sets
$\left(\left\{ x,y,z\right\} ,\left\{ x,y\right\} \right)$ produces
choices $\left(y,x\right)$, both cases reveal that $x$ is preferred
to $y$. The opposite holds for $u\left(y\right)>u\left(x\right)$.
\subsecref{Identification} provides details and an illustration using
\figref{convergence}. \oappxref{supplemental_proofs} \exaref{population}
describes a test using a population of DMs.} This key empirical strategy leverages the fact that an experienced
DM has considered more options and therefore their choices more informatively
convey genuine preference. Moreover, it is possible to identify preferences
using just one carefully designed sequence of choice sets. On the
other hand, attention $\Gamma$ cannot be pinned down even with the
richer dataset of choice sequences, but the set of possible attention
functions can be characterized using a maximal set that allows an
analyst to overestimate or underestimate attention.

The second observations study the link between limited attention and
rationality violations. To see its relevance, note that limited attention
\emph{can} result in ``non-standard'' behavior, but it is not immediately
clear whether it \emph{must}. An investigation into this link begins
with a distinction between two kinds of rationality violations: those
that occur on a \emph{counterfactual} basis and those that will \emph{realize}.\footnote{\subsecref{Counterfactual} provides a formal definition and an illustration
using \figref{primitive}.} Counterfactual violations are typically observed in a between-subject
design, where a population of subjects makes inconsistent decisions
from randomly assigned choice problems; it means some subjects are
predisposed to commit these violations. It turns out that the lack
of counterfactual violations cannot rule out limited attention. A
simple example illustrates a DM who never commits counterfactual violations
but is inconsistent with full attention due to history-dependent behavior.

Unlike counterfactual violations, realized violations come from the
continuous observation of one DM\textemdash when they are seen choosing
$x$ over $y$ at some point and $y$ over $x$ in others. It turns
out that full attention is ruled out if and only if the DM commits
realized WARP violations. If no violations of this kind can be detected,
then the DM is observationally identical to one with full attention.
These observations suggest that realized violation provides the true
test for limited attention even though it is inexorably obscured in
studies that focus on one-shot decisions.

But where does the attention function $\Gamma$ come from? Because
AAT only imposes structure on attention across time, its silence on
attention structures invites a complementary relationship with models
that propose structures for $\Gamma$. \citet{masatlioglu2012revealed}
propose a theory about the intrinsic nature of consideration sets,
that the removal unconsidered alternatives should not affect what
is considered. It can be shown that if the attention function in AAT
satisfies this structure, then future attention, even though it is
evolving as it adapts to past experiences, will continue to satisfy
this structure. The same can be said for the models proposed by \citet{manzini2007sequentially}
and \citet{manzini2012categorize} where criteria like shortlisting
and categorization are used to exclude alternatives from final consideration\textemdash future
attention stems from revised shortlists and re-categorizations. As
a consequence, even though accumulating experience improves decisions,
the DMs do not fundamentally depart from their intrinsic attention
structures.

A complete characterization of compatible models suggests that the
ultimate content of AAT is the correction of WARP violations. One-shot
models that have this feature are compatible with AAT, and it allows
us to connect \emph{attention across time} and \emph{attention across
choice sets}, providing a robust framework of limited consideration
and proposing new ways to test and verify inattention.

Last, the framework is extended to incorporate framing, which enables
an analysis of how different frames can affect a DM's current behavior
and future attention. \textcolor{black}{It begins with a general representation
that captures framing in full generality. Different frames can draw
the DM's attention to different alternatives even though the choice
set is fixed. Successful frames induce lasting consideration for the
future. Special cases of framing are then introduced and characterized,
namely }\textcolor{black}{\emph{ordered lists}}\textcolor{black}{{}
where the DM considers alternatives from top to bottom but may stop
at some point and }\textcolor{black}{\emph{recommendations}}\textcolor{black}{{}
where certain alternatives are made salient. In both cases, postulates
are imposed on choice sequences, proposing new empirical directions
to test whether and how framing works. The model formalizes a number
of intuitive observations, such as the futile repetition of unsuccessful
frames and the crucial role that genuine preference (or quality) plays
in the facilitation of lasting consideration.}

The findings of this paper are undoubtedly limited\textemdash there
is more to learn from choice sequences\textemdash but they underscore
a broader agenda of using richer data, in place of assumptions, to
learn \emph{from} individual behavior. If we believe that behavior
is boundedly rational, then choice sequences emerge as an important
dataset that allows an analyst to observe and study the correction
of \textquotedbl mistakes\textquotedbl . \emph{Whether} corrections
occur, \emph{when} they occur, and \emph{how} they occur each contributes
significantly to a comprehensive examination of bounded rationality.
Limited consideration is only one of many possible examples.

I proceed as follows: Related literature comes next. \secref{Axioms}
introduces the primitive and the axiomatic foundation. \secref{Model}
introduces AAT and basic results regarding identification and rationality
violations. \secref{attention_structures} analyzes different attention
structures in the literature to investigate the link between attention
across time and attention across choice sets. \secref{Frames} extends
the framework to incorporate frames and studies the short- and long-term
effects of framing on choices and attention. \secref{Conclusion}
concludes. \appxref{Proof} contains main proofs. \oappxref{supplemental_proofs}
contains omitted proofs and results.

\subsection{Related Literature}

Closest to AAT is the choice theory literature that studies attention
using consideration sets, even though this literature has not considered
the evolution of choices. \citet{masatlioglu2012revealed}'s \emph{choice
with limited attention} belongs to this category, \citet{lleras2017more}
study consideration sets that preserve considered alternatives in
subsets, and \citet{geng2022limited} introduces triggers and capacity.
Other heuristics and choice procedures can also give rise to consideration
sets, including \citet{manzini2007sequentially}'s \emph{rational
shortlist method}, \citet{manzini2012categorize}'s \emph{categorize-then-choose},
and \citet{cherepanov2013rationalization}'s \emph{rationalization},
\citet{rideout2021choosing}'s \emph{justification} and \citet{geng2021shortlisting}'s\emph{
shortlisting procedure with capacity}, where potentially superior
alternatives are eliminated from final decisions.\footnote{Similar procedures that study the iterative scrutiny of alternatives
can be found in \citet{xu2007rationalizability,apesteguia2013choice,yildiz2016list}} These studies complement AAT's silence on attention structures; in
turn, AAT suggests how choice sequences can serve as a new channel
to study inattention.

Different methods have been deployed to infer preference under limited
attention. \citet{masatlioglu2012revealed}'s CLA already has this
feature, inferring attention partially identifies preference. \citet{caplin2011search2}
and \citet{caplin2011search} study search processes that take place
by observing tentative choices in a single choice problem. \citet{manzini2014stochastic}
and \citet{cattaneo2020random} exploit the richness in stochastic
choice to pin down preference; \citet{kovach2023reference} additionally
consider reference-dependent distributions of consideration sets.
\citet{gossner2021attention} study how behavior may react to the
exogenous manipulation of attention.

The general intuition that past behavior can influence future behavior
is shared elsewhere. In static settings, \citet{gilboa1995case,gilboa2002utility}\emph{
}study\emph{ case-based utility} where a case can be past experiences,
\citet{bordalo2020memory} study how a database of past experiences
can affect the evaluation of alternatives through different perceived
norms. Models of \emph{habit formation} consider DMs who are affected
by the past and attempt to shape future behavior \citep{gul2007harmful,rozen2010foundations,tserenjigmid2020characterization,hayashi2022habit}.

The extension to frames relates to a general setting studied in \citet{salant2008frames}'s
\emph{choice with frames}. \citet{guney2014theory} studies deterministic
behavior under observable lists whereas \citet{manzini2019sequential,tserenjigmid2021order,ishii2021model}
study stochastic behavior/data. \citet{cheung2024disentangling} study
observable recommendation and use stochastic choice data to reveal
the influence of recommendation on both attention and utility. This
paper adds to the literature new ways to test, analyze, and use frames\textemdash using
choice sequences.

\section{\label{sec:Axioms}Setup}

\subsection{Primitive}

Let $X$ be a countable set of alternatives and let $\mathcal{A}$
be the set of all finite subsets of $X$ with at least two elements.\footnote{The exclusion of singleton choice sets is to avoid passive choices,
since ``choosing'' something without a choice might not result in
awareness/consideration. All results go through if we let $\mathcal{A}$
be the set of all finite and nonempty subsets of $X$.} Let $X^{\boldsymbol{\text{N}}}$ be the set of all infinite sequences
of alternatives and let $\mathcal{A}^{\boldsymbol{\text{N}}}$ be
the set of all infinite sequences of choice sets. The primitive is
a choice function that assigns to each infinite sequence of choice
sets an infinite sequence of choices,
\[
c:\mathcal{A}^{\boldsymbol{\text{N}}}\rightarrow X^{\boldsymbol{\text{N}}},
\]
where for every sequence of choice sets $\left(A_{n}\right)\in\mathcal{A}^{\boldsymbol{\text{N}}}$
and any natural number $k$, the corresponding choice $c\left(\left(A_{n}\right)\right)_{k}$
is an element of the corresponding choice set $\left(A_{n}\right)_{k}$.\footnote{Even though I work with infinitely long choice sequences, the main
representation theorem only requires choice sequences to have at least
length 4.} The first choice (when $k=1$) can be treated as the first decision
since the DM enters the analyst's observation. I sometimes write $A_{k}$
for $\left(A_{n}\right)_{k}$. Denote by $\hat{X}$ the set of alternatives
that are \emph{ever-chosen}, i.e., $\hat{X}:=\left\{ x\in X:c\left(\left(A_{n}\right)\right)_{i}=x\text{ for some }\text{\ensuremath{\left(A_{n}\right)} and \ensuremath{i}}\right\} $.\footnote{There is at most one \emph{never-chosen }alternative, since the set
of all binary choice sets leaves at most one alternative never chosen.} Like more standard primitives, this dataset can be extracted from
a population of DMs each randomly assigned to a sequence.\footnote{The concern that a DM can only be observed under one sequence of choice
sets does not prohibit us from discussing what the DM would have chosen
from other choice sequences. The same limitation applies to standard
settings where we consider a DM who produces a choice for each of
multiple choice sets. The solution is to use a between-subject design
with random assignments. If a population of subjects is each randomly
assigned to choice set $A$ or choice set $B$, and aggregate data
presents a WARP violation, then we know that some subjects are predisposed
to commit WARP violations. \oappxref{supplemental_proofs} \exaref{population}
describes this approach for choice sequences.}

To focus on how past choices affect future choices (instead of the
opposite), the scope of this paper is limited to situations in which
two sequences of choice sets that are identical up to a certain point
give the same choices up to that point. This property, henceforth
\emph{future independence}, rules out choices that are made with perfect
foresight of future choice sets. Formally, for any $\left(A_{n}\right),\left(B_{n}\right)\in\mathcal{A}^{\boldsymbol{\text{N}}}$,
if $A_{k}=B_{k}$ for all $k\leq K$, then $c\left(\left(A_{n}\right)\right)_{k}=c\left(\left(B_{n}\right)\right)_{k}$
for all $k\leq K$.

\begin{figure}[h]
\centering{}\includegraphics[scale=0.65]{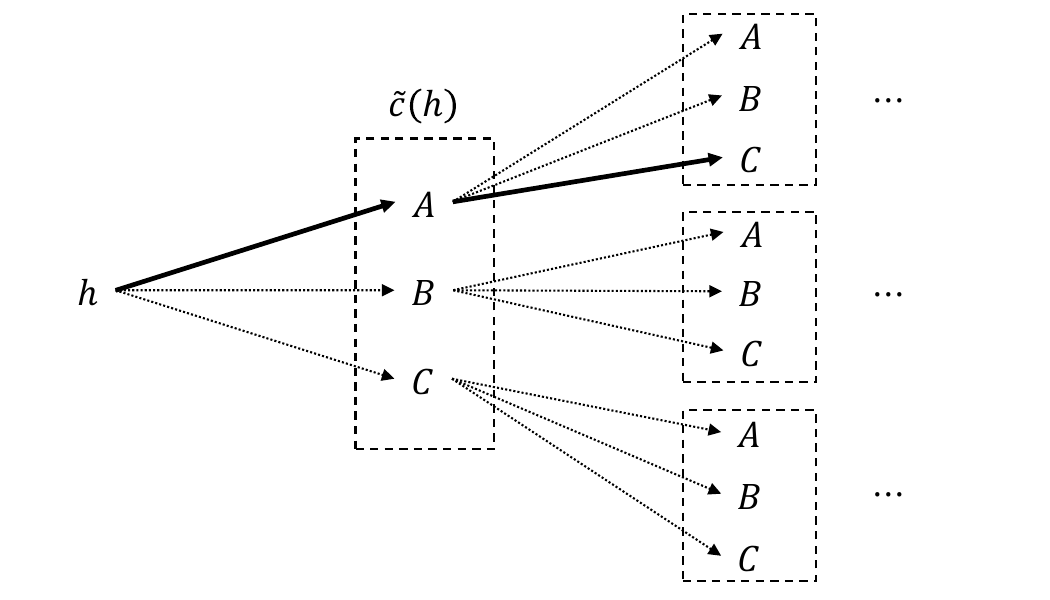}\caption{\label{fig:primitive}Each rectangle is made up of one-shot choice
sets; therefore, a sequence of choice sets includes at most one choice
set from each rectangle. The solid path represents (part of) a sequence
of choice sets.}
\end{figure}

Then, $c$ can be decomposed into an object more familiar to most
economists: a collection of history-dependent choice functions. Formally,
let $\mathcal{A}^{<\boldsymbol{\text{N}}}$ be the set of all finite
sequences of choice sets, including the empty sequence denoted by
$\emptyset$. For each history of choice sets $h\in\mathcal{A}^{<\boldsymbol{\text{N}}}$,
denote by 
\[
\tilde{c}\left(h\right):\mathcal{A}\rightarrow X
\]
the \emph{one-shot choice function} that assigns a choice to each
upcoming choice set (right after $h$). Each $\tilde{c}\left(h\right)$
captures a cross-section of the original primitive $c$, illustrated
in \figref{primitive}. Due to future independence, $\left\{ \tilde{c}\left(h\right)|h\in\mathcal{A}^{<\boldsymbol{\text{N}}}\right\} $
is fully and uniquely pinned down. Choice without (observable) history
is captured by $\tilde{c}\left(\emptyset\right)$, and I refer to
it as $c_{0}$ for convenience.

\subsection{Axioms}

Aligned with the motivation of this paper, all three axioms impose
restrictions on choices across time; they are trivially satisfied
if choice sequences were not considered. The first two axioms provide
basic structure.
\begin{ax}[Weak Stability]
\label{axm:1}For any $\left(A_{n}\right)\in\mathcal{A}^{\boldsymbol{\text{N}}}$
and $h<i<j$, if $c\left(\left(A_{n}\right)\right)_{h}=x$, $c\left(\left(A_{n}\right)\right)_{i}=y\ne x$,
$x\in A_{i}$, and $y\in A_{j}$, then $c\left(\left(A_{n}\right)\right)_{j}\ne x$.
\end{ax}
\axmref{1} imposes a version of the infamous weak axiom of revealed
preferences (WARP) with two key differences.\footnote{There are many (roughly) equivalent definitions for WARP, I use ``if
$x$ is chosen in $y$'s presence, then $y$ is never chosen in $x$'s
presence''.} First, it is imposed \emph{within} a sequence of choice problems
$\left(A_{n}\right)\in\mathcal{A}^{\boldsymbol{\text{N}}}$. In particular,
there is no restriction on how choices differ across sequences.\footnote{Let $A=\left\{ x,y,a\right\} $ and $B=\left\{ x,y,b\right\} $. Suppose
sequence one $\left(A,B,B,B,...\right)$ produces $\left(x,b,b,b,...\right)$
and sequence two $\left(B,A,A,A,...\right)$ produces $\left(y,a,a,a...\right)$,
then choosing $x$ over $y$ in $A$ but the opposite in $B$ is a
WARP violation across sequence, but there is no violation of \axmref{1}.} Second, it does so without demanding full compliance with WARP, but
limits the instances of WARP violations. A conforming DM may switch
between $x$ and $y$, but she does not go back and forth between
them. 

To illustrate, suppose a DM first chooses $x$ in the presence of
$y$ and then chooses $y$ in the presence of $x$. The latter choice
violates WARP, and it may be due to the emerging consideration of
$y$. \axmref{1} does not exclude this behavior, but it posits that
from here on, the choice between $x$ and $y$ has finalized. In other
words, the DM may ``flip'' but must not ``flip-flop''. A DM who
never switches automatically satisfies this axiom.
\begin{ax}[Past Dependence]
\label{axm:2}For any $\left(A_{1},...,A_{K}\right)\in\mathcal{A}^{<\boldsymbol{\text{N}}}$
and $B\in\mathcal{A}$, 
\[
\tilde{c}\left(\left(A_{1},...,A_{K}\right)\right)\left(B\right)\in\left\{ \tilde{c}\left(\left(A_{1},...,A_{K-1}\right)\right)\left(B\right),\tilde{c}\left(\left(A_{1},...,A_{K-1}\right)\right)\left(A_{K}\right)\right\} .
\]
\end{ax}
\axmref{2} allows past choices to affect future choices. One way
to understand this axiom is to first consider the removal of ``$\tilde{c}\left(\left(A_{1},...,A_{K-1}\right)\right)\left(A_{K}\right)$''.
In that case, the DM's choice after history $\left(A_{1},...,A_{K}\right)$
does not depend on whether or not she had experienced $A_{K}$, i.e.,
choices are past \emph{independent}. \axmref{2} weakens past independence
by allowing for one type of departure: the next choice is exactly
the choice it succeeded, $\tilde{c}\left(\left(A_{1},...,A_{K-1}\right)\right)\left(A_{K}\right)$.
That is, after facing a history of choice sets $\left(A_{1},...,A_{K}\right)$,
what a DM chooses from $B$ is either what she would\emph{ }have chosen
had she not experienced $A_{K}$ or exactly what she just chose from
$A_{K}$. The postulate is therefore a delimited weakening of past
independence. First, even though past choices may affect future choices,
it must do so in a period-by-period manner; this provides tractability
and important testable predictions. Moreover, said effect is limited
to ``helping'' the recently chosen alternative to be chosen again;
other forms of past dependence remain prohibited.

The next and last axiom embodies the behavioral signature of attention.
Recall that $c_{0}\left(A\right)$ denotes the choice from $A$ without
(observable) history. Consider a definition of revealed preference
that captures the analyst's inference that $x$ is better than $y$.
This relationship is identified either when $x$ is chosen over $y$
when $y$ was chosen in the past or when $x$ is chosen from a choice
set that $y$ is chosen when there is no observable history. Formally,
for distinct $x$ and $y$, let $xPy$ if at least one of the following
is true for some $\left(A_{n}\right)\in\mathcal{A}^{\boldsymbol{\text{N}}}$:
(1) $c\left(\left(A_{n}\right)\right)_{j}=x$, $y\in A_{j}$ and $c\left(\left(A_{n}\right)\right)_{i}=y$
such that $i<j$ or (2) $c\left(\left(A_{n}\right)\right)_{j}=x$
such that $c_{0}\left(A_{j}\right)=y$.
\begin{ax}[Default Attention]
\label{axm:3}If $c_{0}\left(A\right)Py$, then $y\notin\tilde{c}\left(h\right)\left(A\right)$
for all $h\in\mathcal{A}^{<\boldsymbol{\text{N}}}$.
\end{ax}
\axmref{3} restricts how a DM can depart from her default choice
in a choice set. Specifically, it posits that if the default choice
from $A$, denoted by $c_{0}\left(A\right)$, is identified to be
better than $y$, then $y$ is never going to be chosen from $A$
(no matter the history). Intuitively, it captures the idea that certain
alternatives will always receive attention when they appear in certain
choice sets (regardless of the history), hence \textquotedblleft default
attention\textquotedblright . These may be the most salient alternatives
insofar as to attract attention: pizza is always in the consideration
set for football night, even though the additional consideration of
satay (Malaysian skewers) depends on whether the DM has learned of
this dish. In general, since DM was not born into the analyst's observation,
default attention may be interpreted as the attention structure formed
in the (unobserved) past.

\section{\label{sec:Model}Model}

\subsection{Attention Across Time}

We are ready for the main representation theorem. Denote by $\boldsymbol{c}\left(h\right)$
the set of (previously) chosen alternatives in the history $h$, i.e.,
$\boldsymbol{c}\left(\left(A_{1},...,A_{K}\right)\right):=\left\{ \tilde{c}\left(\emptyset\right)\left(A_{1}\right)\right\} \cup\left\{ \tilde{c}\left(\left(A_{1}\right)\right)\left(A_{2}\right)\right\} \cup\left\{ \tilde{c}\left(\left(A_{1},...,A_{k-1}\right)\right)\left(A_{k}\right):k=3,...,K\right\} .$ 
\begin{defn}
\label{def:AAT}$c$ admits an Attention Across Time (AAT) representation
if there exist a\emph{ utility function} $u:X\rightarrow\mathbb{R}$
and an \emph{attention function} $\Gamma:\mathcal{A}\rightarrow2^{X}\backslash\left\{ \emptyset\right\} $,
where $\Gamma\left(A\right)\subseteq A$, such that
\[
\tilde{c}\left(h\right)\left(A\right)=\underset{x\in\tilde{\Gamma}\left(h\right)\left(A\right)}{\arg\max}\,u\left(x\right)
\]
where $\tilde{\Gamma}\left(h\right)\left(A\right)=\Gamma\left(A\right)\cup\left(\boldsymbol{c}\left(h\right)\cap A\right)$.
\end{defn}
\begin{thm}
\label{thm:AAT}$c$ satisfies Axioms \ref{axm:1}, \ref{axm:2},
and \ref{axm:3} if and only if it admits an Attention Across Time
(AAT) representation.
\end{thm}
AAT suggests the following choice procedure: When a DM faces choice
set $B$, she not only considers alternatives that she would always
consider when she faces $B$ but also the alternatives that she had
chosen in the past. The former is \emph{history-independent} and may
capture what is salient (to her) in the underlying choice problem.
The latter is \emph{history-dependent} and receives her attention
due to her past experiences. The intuition is straightforward\textemdash a
DM may be unaware of certain alternatives, but she must be aware of
the alternatives that she had chosen, which depend on her past experiences.

\paragraph*{Attention structures}

The generality of $\Gamma$ puts no restriction on choice behavior
without history\textemdash every $c_{0}$ is consistent with every
utility function under some $\Gamma$\textemdash but AAT imposes on
what happens \emph{next}.\footnote{For any $c_{0}:\mathcal{A}\rightarrow X$, let $\Gamma\left(A\right)=\left\{ c_{0}\left(A\right)\right\} $
for all $A$ and use any utility function.} For example, consider \citet{masatlioglu2012revealed}'s use of attention
filter to explain why, although $x$ is preferred to $y$, it is chosen
from $\left\{ x,y\right\} $ but not from $\left\{ x,y,z\right\} $.
A possible specification is $\Gamma\left(\left\{ x,y\right\} \right)=\left\{ x,y\right\} $,
$\Gamma\left(\left\{ x,y,z\right\} \right)=\left\{ y,z\right\} $.
AAT accommodates this specification, but it further imposes that if
$x$ is chosen from $\left\{ x,y\right\} $ in the past, then $x$
must be considered when the DM faces $\left\{ x,y,z\right\} $ in
the future, i.e., $x\notin\Gamma\left(\left\{ x,y,z\right\} \right)$
but $x\in\tilde{\Gamma}\left(\left\{ x,y\right\} \right)\left(\left\{ x,y,z\right\} \right)$.
\secref{attention_structures} provides a comprehensive examination
of ``attention across time'' as related to ``attention across choice
sets''.

\paragraph{Inferring preference}

An analyst knows definitively that the DM prefers $a$ to $b$ if
she chose $b$ in the past and chooses $a$ over $b$ in the future,
since both $a$ and $b$ are within the consideration set of the latter
choice problem. In fact, one way to elicit such preference is to first
introduce a choice set under which the DM would choose $b$, and then
ask the DM to choose from a choice set from which she would normally
choose $a$. This simple observation helps pin down preferences even
if attention cannot be directly observed. \subsecref{Identification}
investigates further and proposes empirical strategies to reveal preference.

\paragraph{Variety of experience}

Does having more experience help the DM make better decisions? The
model adds details to this intuition. Better decisions ultimately
come from an expanding $\boldsymbol{c}\left(h\right)$, but facing
the same problem repeatedly does not contribute to the expansion of
$\boldsymbol{c}\left(h\right)$, even though $h$ becomes longer.
Instead, the variety of past experiences is key to expanding $\boldsymbol{c}\left(h\right)$.
To see this, note that if two identical choice sets are faced consecutively,
the choice from the latter must coincide with the choice from the
former, because the consideration set has not expanded. This means
the second decision adds nothing to $\boldsymbol{c}\left(h\right)$.
Similar arguments can be made if a choice set is repeated but not
consecutively. Curiously, increased experience can result in better
decisions that appear ``irrational'', \subsecref{Counterfactual}
provides details.

\paragraph*{History-dependent decisions}

Different alternatives can be chosen from the same choice set at different
points in history, even if the DM is never indifferent and has no
noise in their decisions. Although this qualifies as a (trivial) WARP
violation, their occurrence counter-intuitively suggests that the
DM \emph{used to} be ``irrational,'' because she chose a sub-optimal
option, but has since become more rational through the consideration
of better options. \subsecref{departures} delves deeper into the
significance of history-dependent behaviors.

\paragraph{Relevance of experience}

Certain types of past experiences are more useful than others, and
this depends on what the future entails. To see this, suppose $x$
is added to $\boldsymbol{c}\left(h\right)$ but there is no future
choice set that contains $x$, then the increased awareness of $x$
will not improve future decisions. The same limitation holds if future
choice sets that contain $x$ also contain superior alternatives that
are themselves considered, since $x$ will not be chosen anyway. Therefore,
the relevance of experience holds greater long-term value than sheer
quantity of experience. \secref{Frames} investigates the effects
of framing on future attention; it cautions against using frames to
``help'' a DM because short-term benefits can result in harmful
long-term inattention.

\paragraph*{Other theories}

Choice sequences can invite interest in other theories excluded in
AAT. One possibility is learning, where a DM initially unsure of her
preference discovers that she doesn't like something after consuming
it, causing her to change her behavior in the future. The DM can violate
\axmref{1} by switching back and forth. To see this, consider a DM
in her Penang vacation where she first tries \emph{petai}, realizes
that it is not as good, and then tries \emph{tempoyak} only to find
out that it is worse, sticking to \emph{petai} for the rest of her
trip. Other DMs may be building a bundle to seek variety, or deliberately
randomizing, or simply being stochastic; these behaviors violate \axmref{2}
because the axiom requires the same alternative to be chosen when
a choice set is repeated consecutively.\footnote{Deliberately randomize: \citet{agranov2017stochastic,cerreia2019deliberately}.}

\subsection{\label{subsec:Identification}Identification of Parameters}

Next, a series of observations assert that standard economics problems
that concern welfare and incentives, which rely on the identification
of preferences, are possible to study even when the analyst cannot
directly observe attention.

\subsubsection*{Uniqueness of preference}

The first observation is uniqueness of preferences. In AAT, preference
is unique for ever-chosen alternatives $\hat{X}$. On the contrary,
analyzing past and future choices in isolation often results in a
dilemma trying to conclude whether choices are due to genuine preferences
or due to the lack of attention.
\begin{prop}
\label{prop:uniqueness}If $c$ admits AAT representations $\left(u_{1},\Gamma_{1}\right),\left(u_{2},\Gamma_{2}\right)$
and $x,y\in\hat{X}$, then $u_{1}\left(x\right)>u_{1}\left(y\right)$
if and only if $u_{2}\left(x\right)>u_{2}\left(y\right)$.
\end{prop}

\subsubsection*{Convergence}

To delve deeper, I address the question of \emph{how} to identify
preferences, beginning with an intuitive and useful testable prediction.

\begin{figure}[h]
\centering{}\includegraphics[scale=0.65]{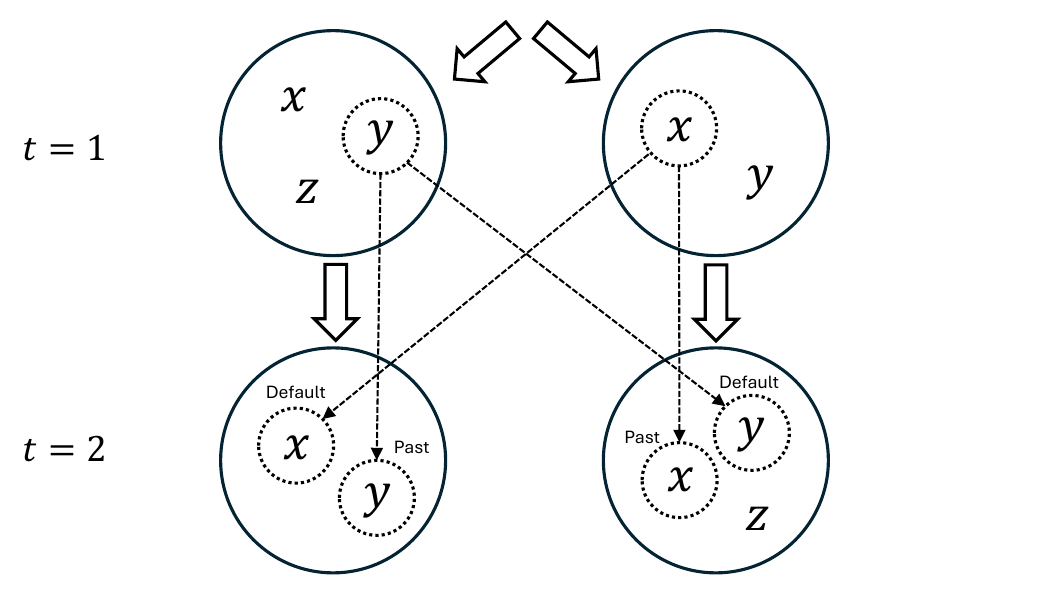}\caption{\label{fig:convergence}Big circles represent choice sets, small circles
represent (inferred) consideration.}
\end{figure}

Consider the following test that compels the correction of ``mistakes''
and reveals preference. Suppose a DM is predisposed to commit a WARP
violation by choosing $x$ from $\left\{ x,y\right\} $ but $y$ from
$\left\{ x,y,z\right\} $. They correspond to the top two circles
in \figref{convergence}, implying that $x$ receives consideration
in $\left\{ x,y\right\} $ and $y$ receives consideration in $\left\{ x,y,z\right\} $,
which is not enough to determine whether $x$ or $y$ is preferred.
However, if a second choice is elicited using alternating choice sets,
then both $x$ and $y$ will receive consideration, one due to default
attention inferred from first period behavior and another due to a
recent experience. The second period choice thus reveals preference.

Moreover, since they maximize the same preference, second period's
choices must coincide, either converging on $x$ or converging on
$y$. This \emph{convergence} property is generalized to arbitrary
choice sets and history in \propref{convergence}; it is an important
testable prediction of AAT and the key indicator that the DM has a
stable preference despite the influence of limited consideration.
\oappxref{supplemental_proofs} \exaref{population} extends this
test to a population of DMs with unknown preference parameters.

\begin{prop}
\label{prop:convergence}Suppose $c$ admits an AAT representation,
$\tilde{c}\left(h\right)\left(A\right)=x$, $\tilde{c}\left(h\right)\left(B\right)=y$,
$\left\{ x,y\right\} \subseteq A\cap B$, and $x\ne y$. Either $\tilde{c}\left(\left(h,A\right)\right)\left(B\right)=\tilde{c}\left(\left(h,B\right)\right)\left(A\right)=x$,
which implies $u\left(x\right)>u\left(y\right)$, or $\tilde{c}\left(\left(h,A\right)\right)\left(B\right)=\tilde{c}\left(\left(h,B\right)\right)\left(A\right)=y$,
which implies $u\left(y\right)>u\left(x\right)$.\footnote{Notation $\left(h,A\right)$ refers to the history that begins with
history $h$ and followed by choice set $A$.}
\end{prop}
Convergence rules out other choice patterns that hold different interpretations,
namely sticky choice (i.e., $\tilde{c}\left(\left\{ x,y\right\} \right)\left(\left\{ x,y,z\right\} \right)=x$,
$\tilde{c}\left(\left\{ x,y,z\right\} \right)\left(\left\{ x,y\right\} \right)=y$),
which may be explained by habit formation, and\emph{ }past independence
(i.e., $\tilde{c}\left(\left\{ x,y\right\} \right)\left(\left\{ x,y,z\right\} \right)=y$,
$\tilde{c}\left(\left\{ x,y,z\right\} \right)\left(\left\{ x,y\right\} \right)=x$),
where it is as if the DM completely neglects past experiences. 

\subsubsection*{Switches}

However, we do not always have the liberty of designing clever tests.
How can we infer preferences in general? Consider a formal definition
of the event that allows us to reveal preference\textemdash switches.
Note that they need not be WARP violations.
\begin{defn}
\label{def:switches}Given $c$ and suppose $x\ne y$.
\begin{enumerate}
\item Let $x\mathbb{S}{}_{\left(A_{n}\right)}y$ if $c\left(\left(A_{n}\right)\right)_{i}=y$
and $c\left(\left(A_{n}\right)\right)_{j}=x$ with $y\in A_{j}$ for
some $i<j$.
\item Let $x\mathbb{S}y$ if $x\mathbb{S}{}_{\left(A_{n}\right)}y$ for
some $\left(A_{n}\right)\in\mathcal{A}^{\boldsymbol{\text{N}}}$.
\end{enumerate}
\end{defn}
Next, \propref{switches} makes a number of observations about switches.
The first is already discussed: if we see a switch from choosing $y$
to choosing $x$ over $y$, then $x$ is preferred. Moreover, this
kind of evidence can always be found as long as both $x$ and $y$
are ever chosen. Also, because preference is unique, evidence can
never be contradictory: if it is possible to reveal that $x$ is preferred
to $y$ in one sequence, it is impossible to reveal the opposite in
any sequence. Finally, if there are only finitely many alternatives,
it is possible to collect all of these evidences using just one sequence
of choice sets by asking the right questions in the right order.\footnote{It is impossible to construct a $c$-independent sequence that fully
identifies preferences. \oappxref{supplemental_proofs} \exaref{impossible_c-independent}
provides a counterexample, \corref{singlesequence_c-indpendent} outlines
the best a $c$-independent sequence can do.}
\begin{prop}
\label{prop:switches}Suppose $c$ admits an AAT representation $\left(u,\Gamma\right)$.
\begin{enumerate}
\item If $x\mathbb{S}y$, then u$\left(x\right)>u\left(y\right)$.
\item The relation $\mathbb{S}$ on $\hat{X}$ is a strict total order.\footnote{A strict total order is a binary relation that is asymmetric (if $x\mathbb{S}y$,
then not $y\mathbb{S}x$), transitive (if $x\mathbb{S}y$ and $y\mathbb{S}z$,
then $x\mathbb{S}z$), and connected (if $x\ne y$, then $x\mathbb{S}y$
or $y\mathbb{S}x$). }
\item If $X$ is finite, then there exists $\left(A_{n}\right)$ such that
$\mathbb{S}{}_{\left(A_{n}\right)}$ on $\hat{X}$ is a strict total
order.
\end{enumerate}
\end{prop}

\subsubsection*{Identification of Attention}

It remains to address whether attention function $\Gamma$ is also
unique, and the answer is negative.

For the behavior that results from full attention, it makes no difference
whether the DM has considered the alternatives that she did not chose;
they are inferior anyway. But the same intuition applies more generally,
since we can always alter $\Gamma$ by adding or removing inferior
alternatives without changing the model's prediction. Therefore, the
set of possible model specifications is characterized by a maximal
set: Given $c$, consider $\Gamma^{+}:\mathcal{A}\rightarrow2^{X}\backslash\left\{ \emptyset\right\} $
where
\[
\Gamma^{+}\left(A\right):=\left\{ x\in A:c_{0}\left(A\right)\mathbb{S}x\text{ or }x=c_{0}\left(A\right)\text{ or }x\notin\hat{X}\right\} .
\]

\begin{prop}
\label{prop:GammaPlus}If $c$ admits an AAT representation, then
it also admits an AAT representation with attention function $\text{\ensuremath{\Gamma}}$
if and only if $c_{0}\left(A\right)\in\Gamma\left(A\right)\subseteq\Gamma^{+}\left(A\right)$
for all $A$.
\end{prop}
In practice, this means an analyst in doubt has the freedom to different
model specifications that would not alter predictions, including overestimating
the size of attention function by taking the union of possible candidates
of $\Gamma$ or by taking a conservative approach using intersections
(\oappxref{supplemental_proofs} \corref{notuniqueattention}). 

\subsection{\label{subsec:Counterfactual}Counterfactual Violations}

Next, I argue that our existing tests for limited attention cannot
possibly be complete without taking choice sequences into account.
The key idea lies in the distinction between two kinds of WARP violations
that are manifested differently: counterfactual and realized.

For every history $h$, $\tilde{c}\left(h\right):\mathcal{A\rightarrow}X$
is a one-shot choice function that captures, at a given point in time,
what the DM would choose for each upcoming choice set. When $\tilde{c}\left(h\right)$
violates WARP, for instance
\[
\tilde{c}\left(h\right)\left(\left\{ x,y\right\} \right)=x\text{ \& }\tilde{c}\left(h\right)\left(\left\{ x,y,z\right\} \right)=y,
\]
it is a \emph{counterfactual} violation. This is typically observed
in a between-subject design where each DM is randomly assigned to
one of two choice sets and aggregate behavior suggests someone is
``non-standard''. It differs from WARP violations that have \emph{realized}
as we observe the same DM over time, obscured in studies that focus
on one-shot decisions, for instance
\[
\tilde{c}\left(\emptyset\right)\left(\left\{ x,y\right\} \right)=x\text{ \& }\tilde{c}\left(\left(\left\{ x,y\right\} \right)\right)\left(\left\{ x,y,z\right\} \right)=y.
\]

Consider \figref{primitive} and suppose $x,y\in A\cap C$. Suppose
in the largest rectangle $x$ is chosen from $A$ and $y$ is chosen
from $C$, then this is a counterfactual WARP violation at history
$h$. In contrast, if along the solid path $x$ is chosen from $A$
and $y$ is chosen from $C$, then this is a \emph{realize}d WARP
violation.

Perhaps unexpected at first, it can be shown that even though full
attention rules out counterfactual WARP violations, the absence of
counterfactual WARP violations does not rule out limited attention.

To see this, \exaref{1} asserts that full attention rules out counterfactual
WARP violations. However, \exaref{2} shows that a DM who satisfies
WARP initially may violate WARP in the future, so the conventional
setting that only considers one-shot decisions can fall short of detecting
all instances of limited attention. Curious at first, it also highlights
that WARP compliance can worsen with increased experience, even though
decisions have definitely improved. Perhaps surprisingly, a DM who\emph{
never} commits counterfactual WARP violations can be (non-trivially)
affected by limited attention, illustrated in \exaref{3}.
\begin{example}
\label{exa:1}Consider an AAT representation where $\Gamma\left(A\right)=A$
for all $A$, so $\text{\ensuremath{\tilde{\Gamma}\left(h\right)\left(A\right)=A}}$
for all $h$, which means the DM considers everything all the time
and therefore by utility maximization, she never commits counterfactual
WARP violations.
\end{example}
\begin{example}
\label{exa:2}Consider $X=\left\{ x,y,z,z'\right\} $ and $\Gamma\left(A\right)=\left\{ z\right\} $
if $z\in A$, $\Gamma\left(A\right)=\left\{ x\right\} $ if $z\notin A$
and $x\in A$, $\Gamma\left(\left\{ y,z'\right\} \right)=\left\{ y\right\} $,
and $u\left(x\right)>u\left(y\right)>u\left(z\right)>u\left(z'\right)$.
Notice that $c_{0}$ can be explained by the maximization of preference
ranking $z\succ_{0}x\succ_{0}y\succ_{0}z'$, thereby complies with
WARP. But with history $h=\left(\left\{ y,z'\right\} \right)$, choices
$\tilde{c}\left(h\right)\left(\left\{ x,y,z,z'\right\} \right)=y$
and $\tilde{c}\left(h\right)\left(\left\{ x,y\right\} \right)=x$
form a counterfactual WARP violation.
\end{example}
\begin{example}
\label{exa:3}Consider $X=\left\{ x,y,z\right\} $ and $\Gamma\left(A\right)=\left\{ x\right\} $
if $x\in A$, $\Gamma\left(\left\{ y,z\right\} \right)=\left\{ y\right\} $,
and $u\left(z\right)>u\left(y\right)>u\left(x\right)$. Notice that
$c_{0}\left(\left\{ x,y\right\} \right)=x$ but $\tilde{c}\left(\left\{ y,z\right\} \right)\left(\left\{ x,y\right\} \right)=y$,
i.e., the choice from $\left\{ x,y\right\} $ varies with history,
so the DM is inconsistent with full attention. However, $c_{0}$ can
be explained by the maximization of preference ranking $x\succ_{0}y\succ_{0}z$,
and it can be shown that the DM never commits counterfactual WARP
violations.\footnote{A WARP violation requires a better alternative to receive attention
in some but not all choice sets, which is impossible when only the
worst alternative of each choice set receives attention initially.
Formally, a WARP violation at $\tilde{c}\left(h\right)$ requires
$a,b\in X$ and $A,B\in\mathcal{A}$ such that $\left\{ a,b\right\} \subseteq A\cap B$,
$u\left(a\right)>u\left(b\right)$, $a\in\tilde{\Gamma}\left(h\right)\left(A\right)$
and $a\notin\tilde{\Gamma}\left(h\right)\left(B\right)$. But this
is not possible. If $a\in\Gamma\left(A\right)$, then it is not possible
that $b\in A$ since only the worst outcome is considered by default.
If $a\notin\Gamma\left(A\right)$, then $a\in\tilde{\Gamma}\left(h\right)\left(A\right)$
implies $a\in\boldsymbol{c}\left(h\right)$, which means $a\in\tilde{\Gamma}\left(h\right)\left(B\right)$,
a contradiction.}
\end{example}
\oappxref{supplemental_proofs} \corref{counterfactual-warp} shows
that a counterfactual WARP violation is present if and only if there
is a sufficiently large difference between initial behavior and true
preference.

\subsection{\label{subsec:departures}Necessary Departures}

If even the lack of counterfactual violations cannot rule out full
attention, can anything rule out full attention? It turns out that
the necessary traces of limited attention lie in realized violations
and history-dependent choices.

Recall that AAT admits the special case where choice behavior is consistent
with full attention, i.e., $\Gamma\left(A\right)=A$ for all $A$.
In this case, every decision appears to be independent of history,
and no WARP violations will be detected, be it counterfactual or realized.
To see this, notice that for all $h$ and $A$,
\[
\tilde{c}\left(h\right)\left(A\right)=\underset{x\in\Gamma\left(A\right)\cup\left(\boldsymbol{c}\left(h\right)\cap A\right)}{\arg\max}\,u\left(x\right)=\underset{x\in A\cup\left(\boldsymbol{c}\left(h\right)\cap A\right)}{\arg\max}\,u\left(x\right)=\underset{x\in A}{\arg\max}\,u\left(x\right).
\]

Of course, just because behavior \emph{can} be represented this way
does not mean the DM actually has full attention; perhaps she got
lucky by only paying attention to the best things.\emph{ }Aware of
this distinction, we avoid calling this behavior full attention and
instead say that it admits a \emph{standard utility representation.}

It turns out that in AAT, behavior must admit a standard utility representation
unless it involves \emph{both} realized WARP violations and history
dependence, making them the quintessential markings of limited consideration.

In order to make this argument formal, consider two new definitions
that result from strengthening \axmref{1} and \axmref{2}.
\begin{defn}
~
\begin{enumerate}
\item \emph{Full Stability:} for every $\left(A_{n}\right)\in\mathcal{A}^{\boldsymbol{\text{N}}}$,
if $c\left(\left(A_{n}\right)\right)_{i}=x$, $\left\{ x,y\right\} \subseteq\left(A_{n}\right)_{i}\cap\left(A_{n}\right)_{j}$,
and $x\ne y$, then $c\left(\left(A_{n}\right)\right)_{j}\ne y$. 
\item \emph{Past Independence:} for every $\left(A_{1},...,A_{K}\right)\in\mathcal{A}^{<\boldsymbol{\text{N}}}$
and $B\in\mathcal{A}$, $\tilde{c}\left(\left(A_{1},...,A_{K}\right)\right)\left(B\right)=\tilde{c}\left(\left(A_{1},...,A_{K-1}\right)\right)\left(B\right)$. 
\end{enumerate}
\end{defn}
Full Stability captures WARP within sequence. Past Independence captures
choices that do vary with past experiences. The former strengthens
\axmref{1} and the latter strengthens \axmref{2}.
\begin{thm}
\label{thm:norelax}Suppose $c$ admits an AAT representation. The
following are equivalent:
\begin{enumerate}
\item $c$ satisfies Full Stability
\item $c$ satisfies Past Independence
\item $c$ admits a standard utility representation
\end{enumerate}
\end{thm}
\thmref{norelax} summarizes the observations we discussed with one
additional finding: Full Stability and Past Independence are linked
under AAT, even though they are in general non-nested and could be
interpreted as different aspects of ``rational'' behaviors. In particular,
Past Independence has remarkable implications: it unites one-shot
choices and choice sequences; if we return to \figref{primitive},
it means the same alternative has to be chosen from $A$ no matter
when and where $A$ appears. If we believe in Past Independence, then
choice sequences are redundant. But bounded rationality often includes
the prospect of correcting ``mistakes'', so Past Independence will
likely fail and choice sequences become important.

Combining \thmref{norelax} and the examples in \subsecref{Counterfactual}
makes the case for studying choice sequences: The lack of counterfactual
violations neither rules out limited attention nor rules out realized
violations, but the lack of realized violations implies consistency
with standard utility representation (\thmref{norelax}) and therefore
rules out counterfactual violations.

\section{\label{sec:attention_structures}Attention Structures}

AAT puts no restriction on attention structures, i.e., $\Gamma$ can
be anything, but an extensive literature\emph{ }suggests that certain
attention structures make more sense than others. For instance, \citet{masatlioglu2012revealed}'s
\emph{choice with limited attention} (CLA) proposes that dropping
an alternative that does not receive consideration should not alter
the consideration set, and hence a particular structure is required
on $\Gamma$. \citet{manzini2007sequentially}'s \emph{rational shortlist
method} (RSM) involves a criterion that removes alternatives from
final consideration and the universal application of the same criterion
imposes structure on $\Gamma$.

It is easy to just plug in these structures into $\Gamma$, but an
immediate concern arises: Will the accumulation of experiences, which
can alter consideration sets, cause a DM to depart from a particular
attention structure? Surprisingly, the answer is probably not.

\subsecref{cla} considers a ``vertical merger'' between AAT and
CLA, \subsecref{rsm} does the same for RSM, and \subsecref{CTC}
does the same for \citet{manzini2012categorize}'s \emph{categorize-then-choose}
(CTC). These models turn out to be highly compatible with AAT; they
introduce structures on attention \emph{across choice sets}, whereas
AAT introduces structure on attention\emph{ across time}. Their complementarity
provides a robust framework and proposes new ways to verify inattention.
\subsecref{all-compatible-models} concludes with a complete characterization
of compatible models.

\subsection{\label{subsec:cla}Attention Filter}

Due to \citet{masatlioglu2012revealed}:
\begin{defn}
\label{def:attention-filter}A mapping $\hat{\Gamma}:\mathcal{A}\rightarrow\mathcal{A}$
is an \emph{attention filter} if $\hat{\Gamma}\left(A\right)\subseteq A$
and $y\notin\hat{\Gamma}\left(A\right)$ implies $\hat{\Gamma}\left(A\backslash\left\{ y\right\} \right)=\hat{\Gamma}\left(A\right)$
\end{defn}
\begin{defn}
\label{def:CLA}A choice function $\hat{c}:\mathcal{A}\rightarrow X$
is a \emph{choice with limited attention} (CLA) if there exist $\hat{u}:X\rightarrow\mathbb{R}$
and an \emph{attention filter} $\hat{\Gamma}:\mathcal{A}\rightarrow\mathcal{A}$
such that 
\[
\hat{c}\left(A\right)=\underset{x\in\bar{\Gamma}\left(x\right)}{\arg\max}\,\hat{u}\left(x\right).
\]
\end{defn}
It turns out that in AAT, a DM who has an attention filter will always
have an (possibly different) attention filter, even though her consideration
sets have changed as she accumulates experiences. \propref{LocalCLA}
formalizes this statement and \exaref{secretmenu} illustrates.
\begin{prop}
\label{prop:LocalCLA}If $c$ admits an AAT representation $\left(u,\Gamma\right)$
where $\Gamma$ is an attention filter, then for any history $h\in\mathcal{A}^{<\boldsymbol{\text{N}}}$,
\begin{enumerate}
\item $\tilde{\Gamma}\left(h\right):\mathcal{A}\rightarrow\mathcal{A}$
is an attention filter,
\item $\tilde{c}\left(h\right):\mathcal{A}\rightarrow X$ is a CLA.
\end{enumerate}
\end{prop}
\begin{example}[Secret Menu]
\label{exa:secretmenu}A fast food chain offers four items, cheeseburger
$a$, hamburger $b$, Flying Dutchman $d$, and Animal Fries $e$.
A customer is initially unaware of the latter two. If both $a$ and
$b$ are unavailable, the chain will recommend $e$, bringing it to
the consumer's attention. And if $e$ is also unavailable, then the
store recommends $d$. The attention function is therefore $\Gamma\left(A\right)=\left\{ a,b\right\} \cap A$
if $\left\{ a,b\right\} \cap A\ne\emptyset$, $\Gamma\left(\left\{ d,e\right\} \right)=\Gamma\left(\left\{ e\right\} \right)=\left\{ e\right\} $,
and $\Gamma\left(\left\{ d\right\} \right)=\left\{ d\right\} $, which
satisfies the property of an attention filter. Suppose $u\left(d\right)>u\left(e\right)>u\left(b\right)>u\left(a\right)$.
Consider the history $h=\left(\left\{ d,e\right\} \right)$ from which
$e$ is considered and chosen, the consumer has since discovered $e$
and includes it in her future consideration sets, i.e., $\tilde{\Gamma}\left(h\right)\left(A\right)=\Gamma\left(A\right)\cup\left\{ e\right\} $
if $e\in A$ and $\tilde{\Gamma}\left(h\right)\left(A\right)=\Gamma\left(A\right)$
otherwise. Although $\tilde{\Gamma}\left(h\right)$ differs from $\Gamma$,
it is still an attention filter because dropping $d$ from a choice
set does not change the consideration set.
\end{example}
One way to characterize these behaviors is to impose CLA's original
axioms on $c_{0}$; they amount to putting (testable) restrictions
on counterfactual choices.\footnote{\citet{masatlioglu2012revealed} proposes an axiom called \emph{WARP
with Limited Attention}.} One might wonder if it is possible to, instead, impose restrictions
on choices across time, and the answer is positive. \axmref{CLA}
introduces the behavioral manifestation of CLA in choice sequences.
It posits that if dropping $y$ results in $x$ no longer chosen\textemdash for
which CLA would infer $x$ is preferred to $y$\textemdash then the
DM should never switch from choosing $x$ to choosing $y$ over $x$.
\begin{ax}
\label{axm:CLA}If $c_{0}\left(T\right)=x$ and $c_{0}\left(T\backslash\left\{ y\right\} \right)\ne x$,
then not $y\mathbb{S}x$.
\end{ax}
\begin{prop}
\label{prop:CLA}$c$ satisfies Axioms \ref{axm:1}, \ref{axm:2},
\ref{axm:3} and \ref{axm:CLA} if and only if it admits an Attention
Across Time (AAT) representation $\text{\ensuremath{\left(u,\Gamma\right)}}$
where $\Gamma$ is an attention filter.
\end{prop}
This robust framework strengthens our identification of parameters.
Results from \secref{Model} carry over, so preferences are pinned
down even though CLA by itself is insufficient. And in the instances
where CLA infers $x$ is preferred to $y$, then the ``direct evidence''
where the DM switches from choosing $y$ to choosing $x$ over $y$
can be found in choice sequences.\footnote{If $c_{0}\left(T\right)=x$, $c_{0}\left(T\backslash\left\{ y\right\} \right)\ne x$,
and $y\in\hat{X}$, then not $y\mathbb{S}x$, so $x\mathbb{S}y$ due
to \propref{switches} (2). The reverse is not always true: if the
DM has $u\left(x\right)>u\left(y\right)$, always considers everything,
and $y$ is not the worst alternative, then $x\mathbb{S}y$ but choices
never violate counterfactual WARP.} On the other hand, CLA narrows down the permissible set of attention
functions to those that are supported by the theory of attention filters,
mitigating the issue of multiplicity and providing important interpretations.

\subsection{\label{subsec:rsm}Shortlisting}

Due to \citet{manzini2007sequentially}:
\begin{defn}
\label{def:RSM}A choice function $\hat{c}:\mathcal{A}\rightarrow X$
is a \emph{rational shortlist methods} (RSM) if there exist asymmetric
binary relations $P_{1}$ and $P_{2}$ on $X$ such that 
\[
\hat{c}\left(A\right)=\max\left(\max\left(A,P_{1}\right),P_{2}\right).
\]
\end{defn}
Here, $\max\left(A,S\right):=\left\{ x\in A|\text{ not }ySx\,\forall y\in A\right\} $.
The model describes a choice procedure that involves sequentially
making a choice, where the DM first creates a shortlist using a rationale,
$P_{1}$, and then makes a final decision using $P_{2}$. The shortlist
can therefore be viewed as a consideration set with certain features,
defined in \defref{shortlistgamma}. Like the case for an attention
filter, \propref{LocalRSM} suggests that a DM who uses a shortlist
will always use a shortlist, even if accumulating experience has changed
the DM's behavior.
\begin{defn}
\label{def:shortlistgamma}A mapping $\hat{\Gamma}:\mathcal{A}\rightarrow\mathcal{A}$
is a \emph{shortlist} if there exists an asymmetric binary relation
$S$ on $X$ such that $\hat{\Gamma}\left(A\right)=\max\left(A,S\right)$
for all $A$.
\end{defn}
\begin{prop}
\label{prop:LocalRSM}If $c$ admits an AAT representation $\left(u,\Gamma\right)$
such that $\Gamma$ is a shortlist, then for any history $h\in\mathcal{A}^{<\boldsymbol{\text{N}}}$,
\begin{enumerate}
\item $\tilde{\Gamma}\left(h\right):\mathcal{A}\rightarrow\mathcal{A}$
is a shortlist,
\item $\tilde{c}\left(h\right):\mathcal{A}\rightarrow X$ is an RSM.
\end{enumerate}
\end{prop}
The merger between RSM and AAT captures an intuitive process: The
DM shortlists alternatives before making final decisions but revises
her rationales as she accumulates experience. Specifically, when a
DM gains experience with an alternative, the original rationale $P_{1}$
is revised so that nothing eliminates said alternative, thereby guaranteeing
its consideration in the future and resulting in better decisions.
\exaref{supplier} illustrates.
\begin{example}[Shortlisting Suppliers]
\label{exa:supplier}A firm can choose from a set of suppliers, but
a Malaysian supplier $a$ and a Thailand supplier $b$ are removed
from consideration when a China supplier $d$ is available. So $\Gamma$
is a shortlist derived from the rationale $bP_{1}a$ and $bP_{1}d$.
Suppose $u\left(a\right)>u\left(b\right)>u\left(d\right)$. During
COVID-19 lockdowns, $d$ is temporarily unavailable, resulting in
the choice set $\left\{ a,b\right\} $ from which $a$ is chosen.
After this experience ($h=\left(\left\{ a,d\right\} \right)$), the
firm cancels $bP_{1}a$ but maintains $bP_{1}d$; so $\tilde{\Gamma}\left(h\right)$
is still a shortlist even though $\tilde{\Gamma}\left(h\right)\ne\Gamma$.
\end{example}
As a consequence, the DM's decisions not only become increasingly
informative of her true preferences but also explain whether her past
choices were in fact influenced by shortlisting. To see this, consider
a sequence of observations where $x$ was initially chosen over $y$,
but after $y$ became chosen in an incidental choice problem, future
comparisons resolve in favor of $y$. This confirms that the initial
choice of $x$ was driven by a rationale that eliminated $y$ instead
of the reflection of genuine preference.

\subsection{\label{subsec:CTC}Dominated Categories}

Related to the intuition of shortlisting is when an entire category
of options is removed from consideration when another category (not
necessarily better) is present. Due to \citet{manzini2012categorize}:
\begin{defn}
\label{def:CTC}A choice function $\hat{c}:\mathcal{A}\rightarrow X$
is \emph{a categorize-then-choose} (CTC) if there exist asymmetric
binary relations $\succ_{s}$ on $2^{X}\backslash\left\{ \emptyset\right\} $
and $\succ^{*}$ on $X$ such that $\hat{c}\left(A\right)=\max\left(\max^{s}\left(A,\succ^{s}\right),\succ^{*}\right).$
\end{defn}
Here, $\max^{s}\left(A,\succ^{s}\right):=\left\{ x\in A|\nexists R',R''\subseteq A:R''\succ^{s}R'\text{ and }x\in R'\right\} $
and $\max\left(A,\succ^{*}\right)$ is defined in \subsecref{rsm}.
The model describes a choice procedure where alternatives belonging
to dominated categories are eliminated using a shading relation $\succ^{s}$
before final decision is made using $\succ^{*}$. The authors describe
the first stage as coarse maximization, using categories. \defref{catgamma}
provides a formal definition.
\begin{defn}
\label{def:catgamma}A mapping $\hat{\Gamma}:\mathcal{A}\rightarrow\mathcal{A}$
is a \emph{coarse-max} if there exists an asymmetric binary relation
$S$ on $2^{X}\backslash\left\{ \emptyset\right\} $ such that $\hat{\Gamma}\left(A\right)=\max^{s}\left(A,S\right)$
for all $A$.
\end{defn}
Similar to before, \propref{LocalCTC} suggests that a DM who coarse-max
will continue to coarse-max even with accumulating experience. However,
future coarse-max becomes ``finer'' as they involve smaller (dominated)
categories, resulting in larger consideration sets and better decisions
due to increased experience. \exaref{italian} illustrates the idea.
\begin{prop}
\label{prop:LocalCTC}If $c$ admits an AAT representation $\left(u,\Gamma\right)$
such that $\Gamma$ is a coarse-max, then for any history $h\in\mathcal{A}^{<\boldsymbol{\text{N}}}$,
\begin{enumerate}
\item $\tilde{\Gamma}\left(h\right):\mathcal{A}\rightarrow\mathcal{A}$
is a coarse-max,
\item $\tilde{c}\left(h\right):\mathcal{A}\rightarrow X$ is a CTC.
\end{enumerate}
\end{prop}
\begin{example}[Favorite Restaurants]
\label{exa:italian}Consider the example from \citet{manzini2012categorize}
where the availability of \{Italian restaurants\} shades \{Mexican
restaurants\}, excluding the latter category from final decisions.
Imagine a history in which a sole Mexican restaurant was open during
Ferragosto, visited by the consumer, and was subsequently re-categorized
as a special Mexican restaurant (perhaps a favorite) exempted from
shading. The other Mexican restaurants continue to be shaded by Italian
restaurants. In the future, the consumer's genuine preference will
determine whether they return to this Mexican restaurant.
\end{example}

\subsection{\label{subsec:all-compatible-models}Characterizing Compatibility}

CLA, RSM, and CTC are different models that capture different behavior,
but their compatibility with AAT contributes to a robust framework
where AAT identifies \emph{preferences} and these models provide insights
to the intrinsic formation of \emph{(in)attention}. Can other models
be compatible? I now characterize a sufficient and necessary property.

Let $X$ be a countable set of alternatives and let $\mathcal{A}$
be the set of all subsets of $X$. Let $\mathbb{C}_{All}$ be the
collection of all (one-shot) choice functions $\hat{c}:\mathcal{A}\rightarrow X$
such that $\hat{c}\left(A\right)\in A$ for all $A\in\mathcal{A}$.
A subset $\mathbb{C}\subseteq\mathbb{C}_{All}$, which may include
some choice functions and exclude others, can be viewed as the universe
of behaviors explained by a given choice model, or equivalently, those
that satisfy some given axioms. Let $\mathbb{C}_{WARP}$ characterize
the set of all choice functions that satisfy WARP (i.e., $\hat{c}\in\mathbb{C}_{WARP}$
if and only if $\hat{c}\left(T\right)=\hat{c}\left(S\right)$ whenever
$\hat{c}\left(S\right)\in T\subseteq S$). By convention, $f\left(\mathcal{A}\right):=\left\{ f\left(A\right):A\in\mathcal{A}\right\} $.
\begin{defn}
\label{def:compatbility}$\mathbb{C}$ is \emph{compatible with AAT}
if for every $f\in\mathbb{\mathbb{C}}$, there exists an AAT representation
such that $c_{0}=f$ and $\tilde{c}\left(h\right)\in\mathbb{C}$ for
all $h\in\mathcal{A}^{<\boldsymbol{\text{N}}}$.
\end{defn}
\defref{compatbility} captures compatibility in the following sense.
Suppose we observe a DM who suffers from limited attention and whose
``non-standard'' behavior $f$ belongs to a choice model $\mathbb{C}$.
If $f$ is a particularly interesting or important behavior, then
the ability for $\mathbb{C}$ to explain $f$ is good news for $\mathbb{C}$.
However, if the DM is also AAT, then her behavior \emph{tomorrow},
$f'$, could differ from $f$, perhaps because her experience today
causes her to expand her awareness or consideration of alternatives.
The question that compatibility asks is whether $f'$ still belongs
to $\mathbb{C}$, and whether this continues to hold when we observe
$f'',f''',f'''',...$ subsequently. If not, then $\mathbb{C}$ and
AAT are not compatible. Note that for $f'$ to belong to $\mathbb{C}$,
there is a certain ``advantage'' for $\mathbb{C}$ to be large and
include many choice functions; but that comes at a cost. An overly
large $\mathbb{C}$ may include a certain odd behavior $g$ where
$g$ or one of $g',g'',g'''...$ fails to belong to $\mathbb{C}$,
resulting again in incompatibility.

Compatibility is therefore non-trivial.\footnote{\oappxref{supplemental_proofs} \exaref{notcompatible} provides a
$\mathbb{C}$ that is not compatible with AAT.} CLA, RSM, and CTC are compatible with AAT due to a common feature
called WARP-convex.
\begin{defn}
\label{def:WARP-convex}Let $f,g,\kappa\in\mathbb{C}_{All}$.
\begin{enumerate}
\item $g$ is a \emph{$\kappa$-cousin} of $f$ if for some finite $T\subseteq f\left(\mathcal{A}\right)$,
$g\left(A\right)=\kappa\left(\left\{ f\left(A\right)\right\} \cup\left[A\cap T\right]\right)$.
\item $\mathbb{C}$ is \emph{WARP-convex} if for all $f\in\mathbb{\mathbb{C}}$,
there exists $\kappa\in\mathbb{C}_{WARP}$ such that every $\kappa$-cousin
of $f$ is in $\mathbb{\mathbb{C}}$.
\end{enumerate}
\end{defn}
Intuitively, a model (or a set of axioms) is WARP-convex if, whenever
a WARP-violating choice function is predicted by the model, choice
functions that can be derived by reconciling some violations with
a WARP-complying choice function can also be explained by the model.
It captures a model's tolerance to the correction of ``mistakes''.

A $\mathbb{C}$ that only contains WARP-conforming choice functions
(even if $\mathbb{C}\ne\mathbb{C}_{WARP}$) is trivially WARP-convex,
including expected utility, exponential discounting, and generalizations
that preserve WARP. The notion is meaningful when we consider non-WARP
models. It turns out that despite its non-triviality, all of the limited
consideration models I looked at are WARP-convex, including RSM \citep{manzini2007sequentially},
CLA \citep{masatlioglu2012revealed}, CTC \citep{manzini2012categorize},
\emph{rationalization} \citep{cherepanov2013rationalization}, and
\emph{overwhelming choice} \citep{lleras2017more}.
\begin{thm}
\label{thm:compatibility}$\mathbb{C}$ is WARP-convex if and only
if it is compatible with AAT.
\end{thm}
\thmref{compatibility} has two components. First, it suggests that
a (one-shot) choice model is compatible with AAT only if its WARP
violations are \emph{correctable}. This direction is intuitive; since
AAT compels the correction of WARP violations in future choices, a
compatible choice model must tolerate a DM who is in the process of
these corrections.

Perhaps unexpected at first, the opposite is also true. A choice model
is compatible with AAT \emph{as long as }WARP violations are correctable,
which highlights the fact that AAT does nothing more than correcting
WARP violations in a specific way. If no WARP violation is present,
then AAT would accommodate a choice behavior as is; but if violations
are present, then AAT would religiously correct them without introducing
new or different types of WARP violations.\footnote{When a model is compatible, it does not mean $\tilde{\Gamma}\left(h\right)$
will, unlike CLA, RSM, and CTC, satisfy the structures imposed by
these models. To see this, \citet{geng2022limited}'s \emph{limited
consideration model with capacity-$k$} puts a cap on the size of
consideration sets. So $\tilde{\Gamma}\left(h\right)\left(A\right)$
will certainly exceed this capacity when $A$ is large and $h$ is
long. Curiously, their model is compatible with AAT, essentially because
we can always find $\hat{\Gamma}\left(h\right)\ne\tilde{\Gamma}\left(h\right)$,
by removing inferior alternatives from consideration, such that $\hat{\Gamma}\left(h\right)$
predicts the same choices while staying within capacity. The same
observation applies to \citet{geng2021shortlisting}'s \emph{shortlisting
with capacity-$k$}.}

\section{\label{sec:Frames}Attention under Frames}

The analysis readily presents an extended framework that incorporates
\emph{frames}. It allows us to study how frames affect attention (now
and in the future) and, more fundamentally, whether frames work.\footnote{There is a general interest in identifying the effects of frames,
see for example \citet{goldin2020revealed}.} A representation for generic frames is introduced, and later specialized
to the cases where frames are \emph{lists} (the DM searches from top
to bottom but may stop at any point) and \emph{recommendations} (the
DM considers recommended alternatives and possibly more).

\subsection{Generic Frames}

I assume that $\mathcal{\bar{A}}$ is a collection of choice problems
with the typical element $\bar{A}=\left(A,F\right)\in\mathcal{\bar{A}}$.
The primitive is a function that assigns to each infinite sequence
of choice problems (choice sets with frames) an infinite sequence
of choices, $c:\bar{\mathcal{A}}^{\boldsymbol{\text{N}}}\rightarrow X^{\boldsymbol{\text{N}}}$,
and other assumptions are analogous to \secref{Axioms}.\footnote{Formally, let $X$ be a countable set of alternatives and let $\mathcal{A}$
be the set of all finite subsets of $X$ with at least two elements.
Let $\bar{\mathcal{A}}$ be a collection of choice problems that satisfies
$\left\{ A:\left(A,F\right)\in\bar{\mathcal{A}}\right\} =\mathcal{A}$,
that is, every choice set appears at least once, even if not every
frame appears in the dataset. I abuse notation by writing ``$x\in\bar{A}$''
when I mean $x\in A$ where $\bar{A}=\left(A,F\right)$. The primitive
$c:\bar{\mathcal{A}}^{\boldsymbol{\text{N}}}\rightarrow X^{\boldsymbol{\text{N}}}$
satisfies $c\left(\left(\bar{A}_{n}\right)\right)_{k}\in\left(\bar{A}_{n}\right)_{k}$
for each $\left(\bar{A}_{n}\right)\in\bar{\mathcal{A}}^{\boldsymbol{\text{N}}}$
and $k\in\mathbb{N}$. I continue to assume future independence. One-shot
choice functions $\tilde{c}$, choice without history $\text{\ensuremath{c_{0}}}$,
and the revealed preference relation $P$ are defined in the same
way as in \secref{Axioms}. I use $\bar{A}_{k}$ for $\left(\bar{A}_{n}\right)_{k}$
and use $f\left(A,F\right)$ for $f\left(\left(A,F\right)\right)$
where $f$ can be $c_{0}$, $\tilde{c}\left(h\right)$, $\Gamma$,
or $\tilde{\Gamma}\left(h\right)$.} For now, a frame is ``generic'' in the sense that it can represent
any observable difference in the presentation of alternatives. Choices
are therefore captured by a collection of history dependent one-shot
choice functions,
\[
\tilde{c}\left(h\right):\mathcal{\bar{A}}\rightarrow X,
\]
where $h\in\bar{\mathcal{A}}^{\boldsymbol{<\text{N}}}$ is a finite
sequence of choice sets each presented under some frame. Importantly,
the DM can make different choices for the same choice set $A$ when
it appears under different frames, i.e., $\tilde{c}\left(h\right)\left(A,F\right)\ne\tilde{c}\left(h\right)\left(A,F'\right)$.

It turns out that the original set of axioms in \secref{Axioms},
after cosmetic modifications, suffices for an AAT representation with
frames.\footnote{\begin{ax}
\label{axm:frame-1}For any $\left(\bar{A}_{n}\right)\in\mathcal{\bar{A}}^{\boldsymbol{\text{N}}}$
and $h<i<j$, if $c\left(\left(\bar{A}_{n}\right)\right)_{h}=x$,
$c\left(\left(\bar{A}_{n}\right)\right)_{i}=y\ne x$, $x\in\bar{A}_{i}$,
and $y\in\bar{A}_{j}$, then $c\left(\left(\bar{A}_{n}\right)\right)_{j}\ne x$.
\end{ax}
\begin{ax}
\label{axm:frame-2}For any $\bar{B}\in\mathcal{\bar{A}}$, $\tilde{c}\left(\left(\bar{A}_{1},...,\bar{A}_{K}\right)\right)\left(\bar{B}\right)$$\in\left\{ \tilde{c}\left(\left(\bar{A}_{1},...,\bar{A}_{K-1}\right)\right)\left(\bar{B}\right),\tilde{c}\left(\left(\bar{A}_{1},...,\bar{A}_{K-1}\right)\right)\left(\bar{A}_{K}\right)\right\} $.
\end{ax}
\begin{ax}
\label{axm:frame-3}If $c_{0}\left(\bar{A}\right)Py$, then $y\notin\tilde{c}\left(h\right)\left(\bar{A}\right)$
for all $h\in\bar{A}$.
\end{ax}
}
\begin{defn}
\label{def:AATF}$c$ admits an Attention Across Time with Frames
(AAT-F) representation if there exist a\emph{ utility function} $u:X\rightarrow\mathbb{R}$
and an \emph{attention function} $\Gamma:\bar{\mathcal{A}}\rightarrow2^{X}\backslash\left\{ \emptyset\right\} $,
where $\Gamma\left(A,F\right)\subseteq A$, such that
\[
\tilde{c}\left(h\right)\left(A,F\right)=\underset{x\in\tilde{\Gamma}\left(h\right)\left(A,F\right)}{\arg\max}\,u\left(x\right)
\]
where $\tilde{\Gamma}\left(h\right)\left(A,F\right)=\Gamma\left(A,F\right)\cup\left(\boldsymbol{c}\left(h\right)\cap A\right)$.
\end{defn}
\begin{thm}
\label{thm:AATF}$c$ satisfies Axioms \ref{axm:frame-1}, \ref{axm:frame-2},
and \ref{axm:frame-3} if and only if it admits an Attention Across
Time with Frames (AATF) representation.
\end{thm}
As before, the DM pays attention to a (weak) subset of alternatives
$\Gamma\left(A,F\right)\text{\ensuremath{\subseteq A}}$ and considers
historically chosen alternatives $\boldsymbol{c}\left(h\right)$ if
they are available. But unlike before, different frames can induce
different consideration sets for same choice set $A$, i.e., $\Gamma\left(A,F'\right)\ne\Gamma\left(A,F''\right)$,
thereby resulting in different choices. Because these choices will
remain in lasting consideration, frames have short-term and long-term
effects.

\subsection{Effective Frames}

A number of basic observations capture the effect of framing in this
framework. 

First, \exaref{successful} suggests that a frame can successfully
draw the DM's attention to a target alternative now and in the future.
\begin{example}
\label{exa:successful}Suppose the history is $h$ and a certain frame
$F$ is introduced for choice set $A$, with the intention of alerting
the DM to a target alternative $x$. If the DM ends up choosing $x$,
i.e., $\tilde{c}\left(h\right)\left(A,F\right)=x$, then the frame
is successful and the DM will always consider $x$ in the future,
i.e., $x\in\tilde{\Gamma}\left(h'\right)\left(A',F'\right)$ if $x\in A'$,
where $h'$ is any history that begins with $h$ and $\left(A,F\right)$.
\end{example}
However, \exaref{unsuccessful} suggests that repeating an unsuccessful
frame will neither alter behavior now nor affect attention in the
future.
\begin{example}
\label{exa:unsuccessful}In \exaref{successful}, if the frame was
not successful, i.e., $\tilde{c}\left(h\right)\left(A,F\right)\ne x$,
then repeating the same frame will be futile. To see this, suppose
$\tilde{c}\left(h\right)\left(A,F\right)=y$ and let $h'$ be the
history $h$ followed by $\left(A,F\right)$. Since additional consideration
is paid only to the newly chosen alternative $y$, which was already
receiving consideration, the DM's consideration set when she encounters
$\left(A,F\right)$ for the second time is the same as the first time,
i.e., $\tilde{\Gamma}\left(h'\right)\left(A,F\right)=\tilde{\Gamma}\left(h\right)\left(A,F\right)$,
which results in the same choice $\tilde{c}\left(h'\right)\left(A,F\right)=y$.
Since the target alternative $x$ is still not chosen, it will not
be added into future considerations.
\end{example}
These observations capture some intuitive aspects of framing. In particular,
if the target alternative is inferior, then using a frame to elevate
it will not produce meaningful results; the DM will simply consider
it and choose something else. Because lasting consideration can only
result from the DM choosing the target alternative at least once,
the model suggests that effective framing still depends substantially
on the relative quality of the target alternative in the choice set
(or the consideration set). For example, if a frame can cause more
appealing alternatives to \emph{not} receive consideration, then it
may help the target alternative to be chosen now and remain in consideration
in the future.

Can we help a DM by alerting them to superior alternatives? The model
cautions this endeavor. \secref{Model} highlights that the quality
of future decisions depends on the complementary between past experiences
and future problems. For instance, alerting the DM to a better alternative
tends to be effective (since it is better, the DM will choose it)
and improves current utility; but if this alternative is unavailable
in future, then the intervention could lead to negative long-term
consequences where an always-available option never receives consideration,
as in \exaref{hiddentalent}.
\begin{example}[Hidden Talent]
\label{exa:hiddentalent}A stand-up comedy show needs an emergency
substitute, and only fringe performers $a$ (very talented, an economist)
and $b$ (talented, a full-time performer) are available for last-minute
arrangements. The show is only aware of $b$ and would have chosen
her, but a friend brings $a$ to the show's consideration resulting
in the hiring of $a$. However, $a$ later returns to her full-time
job and is no longer available in the future. Facing the normal selection
of performers, the show does not consider $b$, even though $b$ is
better than most of the chosen performers. The show's overall utility
could be better had it chosen $b$ earlier and kept her in lasting
consideration.
\end{example}

\subsection{Ordered Lists}

Suppose each frame is an ordered list, that is, for each $\bar{A}=\left(A,F\right)\in\bar{\mathcal{A}}$,
$F$ is a complete, transitive, and antisymmetric binary relation
(a linear order) on $A$, where $yFx$ ($y\ne x$) is interpreted
as ``$y$ is listed above $x$''.

Do DMs search from the top just because alternatives are ordered this
way? Consider \axmref{frame-list}, which says that if $x$ is chosen
from $\bar{A}=\left(A,F\right)$ when there is no (observable) history,
in which it is listed below $y$ (according to $F$), then a switch
from $x$ to $y$ cannot occur, i.e., not $y\mathbb{S}x$.\footnote{Similar to \secref{Model}, we say $y\mathbb{S}x$ if there exists
$\left(\bar{A}_{n}\right)\in\mathcal{\bar{A}}^{\boldsymbol{\text{N}}}$
such that $c\left(\left(\bar{A}_{n}\right)\right)_{i}=x$ and $c\left(\left(\bar{A}_{n}\right)\right)_{j}=y\ne x$
with $x\in\bar{A}_{j}$ for some $i<j$.} Intuitively, the consideration of $x$ \emph{should} have implied
the consideration of $y$, assuming that the ordered list works, and
choosing $x$ means $x$ is better than $y$; therefore the DM has
no reason to switch from choosing $x$ to choosing $y$ over $x$.
\begin{ax}
\label{axm:frame-list}If $yFc_{0}\left(\bar{A}\right)$ for some
$\bar{A}=\left(A,F\right)$, then not $y\mathbb{S}c_{0}\left(\bar{A}\right)$.
\end{ax}
It turns out that this postulate fully characterizes the expected
behavior for a top to bottom search.
\begin{prop}
\label{prop:frame-list}$c$ satisfies Axioms \ref{axm:frame-1},
\ref{axm:frame-2}, \ref{axm:frame-3} and \ref{axm:frame-list} if
and only if it admits an AATF representation where $x\in\Gamma\left(A,F\right)$
and $yFx$ implies $\text{\ensuremath{y\in}}\Gamma\left(A,F\right)$.
\end{prop}
\begin{example}[Baby Diapers]
An academic goes online to purchase baby diapers for a newborn. The
search result presents 100 brands in a list and the academic searches
from top to bottom but does not consider everything, stopping at a
certain point (say, 15 brands). It is not clear if the academic ultimately
purchases the most preferred diaper, but if item number 13, brand
$x$, is purchased, then it is preferred over the previous 12 items.
Moreover, when the academic returns to top up on diapers, brand $x$
may be listed beyond the point the academic searches, but already
aware of this brand the academic considers it anyway.
\end{example}

\subsection{Recommendations}

Consider one last application where a frame is a set of recommended
options. For each $\bar{A}=\left(A,F\right)\in\bar{\mathcal{A}}$,
$F\left(A\right)\subseteq A$ is a set of alternatives that are highlighted
or made salient to the DM. For each choice set $A$, different recommendations
$F,F'$ can be made, and they potentially result in different decisions.

Whether or not recommendations work, and whether they work as intended,
is neither assumed nor observed; we infer using choice sequences.
\axmref{frame-rec} says that if $y$ is a recommended option in choice
problem $\bar{A}$, but the DM chooses $x$, then a switch from $x$
to $y$ cannot occur.
\begin{ax}
\label{axm:frame-rec}If $y\in F\left(A\right)$ for some $\bar{A}=\left(A,F\right)$,
then not $y\mathbb{S}c_{0}\left(\bar{A}\right)$.
\end{ax}
\begin{prop}
\label{prop:frame-rec}$c$ satisfies Axioms \ref{axm:frame-1}, \ref{axm:frame-2},
\ref{axm:frame-3} and \ref{axm:frame-rec} if and only if it admits
an AATF representation where $x\in F\left(A\right)$ implies $x\in\Gamma\left(A,F\right)$.
\end{prop}
Although consideration sets cannot be directly observed, \axmref{frame-rec}
provides a test for whether recommendations work. If the axiom fails,
that means there is a choice problem $\bar{A}$ from which the decision
maker fails to consider a recommended option. On the contrary, if
the axiom is satisfied, then behavior is consistent with consideration
sets that include all recommended options.
\begin{example}[Unsought Advice]
Out of two libraries on campus, Lehman $a$ and Butler $b$, a professor
recommends $a$ to a PhD student, expecting that the student will
at least consider $a$ (the student has a consistent preference and
knows which one is better as long as it is considered). Surprisingly,
the student chooses $b$, leading the professor to conclude that the
student prefers $b$ over $a$. A couple of months later, the student
stops going to $b$ and switches to $a$, indicating to the professor
that their initial recommendation was, in fact, disregarded.
\end{example}

\section{\label{sec:Conclusion}Conclusion}

This paper introduces a framework that studies how past experiences
can lead to the consideration of previously chosen alternatives in
future decisions. The intuition is captured by a model called \emph{Attention
Across Time} (AAT), which allows an analyst to fully pin down preferences
even if attention is not directly observed, paving the way to sharper
welfare analysis in the presence of limited attention. A wide range
of implications is drawn, including empirical techniques to reveal
preferences, the separation of counterfactual and realized violations,
more robust tests of limited consideration, compatibility with different
attention structures, and an extension for studying the short- and
long-term effects of frames. The findings center around one key message:
that the wealth of information in choice sequences contributes meaningfully
to the examination of boundedly rational behavior, where limited consideration
is one of many possible examples. Should choice sequences receive
lasting consideration?

\begin{singlespace}\bibliographystyle{chicago}
\phantomsection\addcontentsline{toc}{section}{\refname}\bibliography{database_aat}

\begin{thebibliography}{}

\bibitem[\protect\citeauthoryear{Agranov and Ortoleva}{Agranov and
  Ortoleva}{2017}]{agranov2017stochastic}
Agranov, M. and P.~Ortoleva (2017).
\newblock Stochastic choice and preferences for randomization.
\newblock {\em Journal of Political Economy\/}~{\em 125\/}(1), 40--68.

\bibitem[\protect\citeauthoryear{Aguiar}{Aguiar}{2017}]{aguiar2017random}
Aguiar, V. (2017).
\newblock Random categorization and bounded rationality.
\newblock {\em Economics Letters\/}~{\em 159}, 46--52.

\bibitem[\protect\citeauthoryear{Ahn, Iijima, Le~Yaouanq, and Sarver}{Ahn
  et~al.}{2019}]{ahn2019behavioural}
Ahn, D.~S., R.~Iijima, Y.~Le~Yaouanq, and T.~Sarver (2019).
\newblock Behavioural characterizations of naivete for time-inconsistent
  preferences.
\newblock {\em The Review of Economic Studies\/}~{\em 86\/}(6), 2319--2355.

\bibitem[\protect\citeauthoryear{Apesteguia and Ballester}{Apesteguia and
  Ballester}{2013}]{apesteguia2013choice}
Apesteguia, J. and M.~A. Ballester (2013).
\newblock Choice by sequential procedures.
\newblock {\em Games and Economic Behavior\/}~{\em 77\/}(1), 90--99.

\bibitem[\protect\citeauthoryear{Bordalo, Gennaioli, and Shleifer}{Bordalo
  et~al.}{2020}]{bordalo2020memory}
Bordalo, P., N.~Gennaioli, and A.~Shleifer (2020).
\newblock Memory, attention, and choice.
\newblock {\em The Quarterly Journal of Economics\/}~{\em 135\/}(3),
  1399--1442.

\bibitem[\protect\citeauthoryear{Brady and Rehbeck}{Brady and
  Rehbeck}{2016}]{brady2016menu}
Brady, R.~L. and J.~Rehbeck (2016).
\newblock Menu-dependent stochastic feasibility.
\newblock {\em Econometrica\/}~{\em 84\/}(3), 1203--1223.

\bibitem[\protect\citeauthoryear{Caplin and Dean}{Caplin and
  Dean}{2011}]{caplin2011search2}
Caplin, A. and M.~Dean (2011).
\newblock Search, choice, and revealed preference.
\newblock {\em Theoretical Economics\/}~{\em 6\/}(1), 19--48.

\bibitem[\protect\citeauthoryear{Caplin, Dean, and Martin}{Caplin
  et~al.}{2011}]{caplin2011search}
Caplin, A., M.~Dean, and D.~Martin (2011).
\newblock Search and satisficing.
\newblock {\em American Economic Review\/}~{\em 101\/}(7), 2899--2922.

\bibitem[\protect\citeauthoryear{Cattaneo, Ma, Masatlioglu, and
  Suleymanov}{Cattaneo et~al.}{2020}]{cattaneo2020random}
Cattaneo, M.~D., X.~Ma, Y.~Masatlioglu, and E.~Suleymanov (2020).
\newblock A random attention model.
\newblock {\em Journal of Political Economy\/}~{\em 128\/}(7), 2796--2836.

\bibitem[\protect\citeauthoryear{Cerreia-Vioglio, Dillenberger, Ortoleva, and
  Riella}{Cerreia-Vioglio et~al.}{2019}]{cerreia2019deliberately}
Cerreia-Vioglio, S., D.~Dillenberger, P.~Ortoleva, and G.~Riella (2019).
\newblock Deliberately stochastic.
\newblock {\em American Economic Review\/}~{\em 109\/}(7), 2425--2445.

\bibitem[\protect\citeauthoryear{Cherepanov, Feddersen, and
  Sandroni}{Cherepanov et~al.}{2013}]{cherepanov2013rationalization}
Cherepanov, V., T.~Feddersen, and A.~Sandroni (2013).
\newblock Rationalization.
\newblock {\em Theoretical Economics\/}~{\em 8\/}(3), 775--800.

\bibitem[\protect\citeauthoryear{Cheung and Masatlioglu}{Cheung and
  Masatlioglu}{2024}]{cheung2024disentangling}
Cheung, P. and Y.~Masatlioglu (2024).
\newblock Disentangling attention and utility channels in recommendations.
\newblock Working paper.

\bibitem[\protect\citeauthoryear{Chew, Kader, and Wang}{Chew
  et~al.}{2022}]{chew2022source}
Chew, S.~H., G.~Kader, and W.~Wang (2022).
\newblock Source recursive expected utility on rich mixture sets.
\newblock Working paper.

\bibitem[\protect\citeauthoryear{Dean, K{\i}br{\i}s, and Masatlioglu}{Dean
  et~al.}{2017}]{dean2017limited}
Dean, M., {\"O}.~K{\i}br{\i}s, and Y.~Masatlioglu (2017).
\newblock Limited attention and status quo bias.
\newblock {\em Journal of Economic Theory\/}~{\em 169}, 93--127.

\bibitem[\protect\citeauthoryear{Dekel, Lipman, and Rustichini}{Dekel
  et~al.}{2009}]{dekel2009temptation}
Dekel, E., B.~L. Lipman, and A.~Rustichini (2009).
\newblock Temptation-driven preferences.
\newblock {\em The Review of Economic Studies\/}~{\em 76\/}(3), 937--971.

\bibitem[\protect\citeauthoryear{Dillenberger}{Dillenberger}{2010}]{dillenberger2010preferences}
Dillenberger, D. (2010).
\newblock Preferences for one-shot resolution of uncertainty and allais-type
  behavior.
\newblock {\em Econometrica\/}~{\em 78\/}(6), 1973--2004.

\bibitem[\protect\citeauthoryear{Dillenberger and Sadowski}{Dillenberger and
  Sadowski}{2012}]{dillenberger2012ashamed}
Dillenberger, D. and P.~Sadowski (2012).
\newblock Ashamed to be selfish.
\newblock {\em Theoretical Economics\/}~{\em 7\/}(1), 99--124.

\bibitem[\protect\citeauthoryear{Echenique, Saito, and Tserenjigmid}{Echenique
  et~al.}{2018}]{echenique2018perception}
Echenique, F., K.~Saito, and G.~Tserenjigmid (2018).
\newblock The perception-adjusted luce model.
\newblock {\em Mathematical Social Sciences\/}~{\em 93}, 67--76.

\bibitem[\protect\citeauthoryear{Ellis and Masatlioglu}{Ellis and
  Masatlioglu}{2022}]{ellis2022choice}
Ellis, A. and Y.~Masatlioglu (2022).
\newblock Choice with endogenous categorization.
\newblock {\em The Review of Economic Studies\/}~{\em 89\/}(1), 240--278.

\bibitem[\protect\citeauthoryear{Freeman}{Freeman}{2021}]{freeman2021revealing}
Freeman, D.~J. (2021).
\newblock Revealing na{\"\i}vet{\'e} and sophistication from procrastination
  and preproperation.
\newblock {\em American Economic Journal: Microeconomics\/}~{\em 13\/}(2),
  402--38.

\bibitem[\protect\citeauthoryear{Fudenberg, Iijima, and Strzalecki}{Fudenberg
  et~al.}{2015}]{fudenberg2015stochastic}
Fudenberg, D., R.~Iijima, and T.~Strzalecki (2015).
\newblock Stochastic choice and revealed perturbed utility.
\newblock {\em Econometrica\/}~{\em 83\/}(6), 2371--2409.

\bibitem[\protect\citeauthoryear{Geng}{Geng}{2022}]{geng2022limited}
Geng, S. (2022).
\newblock Limited consideration model with a trigger or a capacity.
\newblock {\em Journal of Mathematical Economics\/}~{\em 101}, 102692.

\bibitem[\protect\citeauthoryear{Geng and {\"O}zbay}{Geng and
  {\"O}zbay}{2021}]{geng2021shortlisting}
Geng, S. and E.~Y. {\"O}zbay (2021).
\newblock Shortlisting procedure with a limited capacity.
\newblock {\em Journal of Mathematical Economics\/}~{\em 94}, 102447.

\bibitem[\protect\citeauthoryear{Gilboa and Schmeidler}{Gilboa and
  Schmeidler}{1995}]{gilboa1995case}
Gilboa, I. and D.~Schmeidler (1995).
\newblock Case-based decision theory.
\newblock {\em The Quarterly Journal of Economics\/}~{\em 110\/}(3), 605--639.

\bibitem[\protect\citeauthoryear{Gilboa, Schmeidler, and Wakker}{Gilboa
  et~al.}{2002}]{gilboa2002utility}
Gilboa, I., D.~Schmeidler, and P.~P. Wakker (2002).
\newblock Utility in case-based decision theory.
\newblock {\em Journal of Economic Theory\/}~{\em 105\/}(2), 483--502.

\bibitem[\protect\citeauthoryear{Goldin and Reck}{Goldin and
  Reck}{2020}]{goldin2020revealed}
Goldin, J. and D.~Reck (2020).
\newblock Revealed-preference analysis with framing effects.
\newblock {\em Journal of Political Economy\/}~{\em 128\/}(7), 2759--2795.

\bibitem[\protect\citeauthoryear{Gossner, Steiner, and Stewart}{Gossner
  et~al.}{2021}]{gossner2021attention}
Gossner, O., J.~Steiner, and C.~Stewart (2021).
\newblock Attention please!
\newblock {\em Econometrica\/}~{\em 89\/}(4), 1717--1751.

\bibitem[\protect\citeauthoryear{Gul, Natenzon, and Pesendorfer}{Gul
  et~al.}{2014}]{gul2014random}
Gul, F., P.~Natenzon, and W.~Pesendorfer (2014).
\newblock Random choice as behavioral optimization.
\newblock {\em Econometrica\/}~{\em 82\/}(5), 1873--1912.

\bibitem[\protect\citeauthoryear{Gul and Pesendorfer}{Gul and
  Pesendorfer}{2001}]{gul2001temptation}
Gul, F. and W.~Pesendorfer (2001).
\newblock Temptation and self-control.
\newblock {\em Econometrica\/}~{\em 69\/}(6), 1403--1435.

\bibitem[\protect\citeauthoryear{Gul and Pesendorfer}{Gul and
  Pesendorfer}{2006}]{gul2006random}
Gul, F. and W.~Pesendorfer (2006).
\newblock Random expected utility.
\newblock {\em Econometrica\/}~{\em 74\/}(1), 121--146.

\bibitem[\protect\citeauthoryear{Gul and Pesendorfer}{Gul and
  Pesendorfer}{2007}]{gul2007harmful}
Gul, F. and W.~Pesendorfer (2007).
\newblock Harmful addiction.
\newblock {\em The Review of Economic Studies\/}~{\em 74\/}(1), 147--172.

\bibitem[\protect\citeauthoryear{Guney}{Guney}{2014}]{guney2014theory}
Guney, B. (2014).
\newblock A theory of iterative choice in lists.
\newblock {\em Journal of Mathematical Economics\/}~{\em 53}, 26--32.

\bibitem[\protect\citeauthoryear{Halevy and Ozdenoren}{Halevy and
  Ozdenoren}{2022}]{halevy2022uncertainty}
Halevy, Y. and E.~Ozdenoren (2022).
\newblock Uncertainty and compound lotteries: Calibration.
\newblock {\em Economic Theory\/}~{\em 74\/}(2), 373--395.

\bibitem[\protect\citeauthoryear{Hauser and Wernerfelt}{Hauser and
  Wernerfelt}{1990}]{hauser1990evaluation}
Hauser, J.~R. and B.~Wernerfelt (1990).
\newblock An evaluation cost model of consideration sets.
\newblock {\em Journal of Consumer Research\/}~{\em 16\/}(4), 393--408.

\bibitem[\protect\citeauthoryear{Hayashi and Takeoka}{Hayashi and
  Takeoka}{2022}]{hayashi2022habit}
Hayashi, T. and N.~Takeoka (2022).
\newblock Habit formation, self-deception, and self-control.
\newblock {\em Economic Theory\/}~{\em 74\/}(2), 547--592.

\bibitem[\protect\citeauthoryear{Ishii, Kovach, and {\"U}lk{\"u}}{Ishii
  et~al.}{2021}]{ishii2021model}
Ishii, Y., M.~Kovach, and L.~{\"U}lk{\"u} (2021).
\newblock A model of stochastic choice from lists.
\newblock {\em Journal of Mathematical Economics\/}~{\em 96}, 102509.

\bibitem[\protect\citeauthoryear{Kahneman and Tversky}{Kahneman and
  Tversky}{1979}]{kahneman1979prospect}
Kahneman, D. and A.~Tversky (1979).
\newblock Prospect theory: An analysis of choice under risk.
\newblock {\em Econometrica\/}~{\em 47\/}(2), 263--292.

\bibitem[\protect\citeauthoryear{Ke and Zhang}{Ke and
  Zhang}{2023}]{ke2023multidimensional}
Ke, S. and M.~Zhang (2023).
\newblock Multidimensional choices under uncertainty.
\newblock Working paper.

\bibitem[\protect\citeauthoryear{Kibris, Masatlioglu, and Suleymanov}{Kibris
  et~al.}{2024}]{kibris2024random}
Kibris, Ã., Y.~Masatlioglu, and E.~Suleymanov (2024).
\newblock A random reference model.
\newblock {\em American Economic Journal: Microeconomics\/}~{\em 16\/}(1),
  155--209.

\bibitem[\protect\citeauthoryear{Kovach}{Kovach}{2020}]{kovach2020twisting}
Kovach, M. (2020).
\newblock Twisting the truth: Foundations of wishful thinking.
\newblock {\em Theoretical Economics\/}~{\em 15\/}(3), 989--1022.

\bibitem[\protect\citeauthoryear{Kovach and Suleymanov}{Kovach and
  Suleymanov}{2023}]{kovach2023reference}
Kovach, M. and E.~Suleymanov (2023).
\newblock Reference dependence and random attention.
\newblock {\em Journal of Economic Behavior \& Organization\/}~{\em 215},
  421--441.

\bibitem[\protect\citeauthoryear{Kovach and Tserenjigmid}{Kovach and
  Tserenjigmid}{2022}]{kovach2022behavioral}
Kovach, M. and G.~Tserenjigmid (2022).
\newblock Behavioral foundations of nested stochastic choice and nested logit.
\newblock {\em Journal of Political Economy\/}~{\em 130\/}(9), 2411--2461.

\bibitem[\protect\citeauthoryear{Lanzani}{Lanzani}{2022}]{lanzani2022correlation}
Lanzani, G. (2022).
\newblock Correlation made simple: Applications to salience and regret theory.
\newblock {\em The Quarterly Journal of Economics\/}~{\em 137\/}(2), 959--987.

\bibitem[\protect\citeauthoryear{Lleras, Masatlioglu, Nakajima, and
  Ozbay}{Lleras et~al.}{2017}]{lleras2017more}
Lleras, J.~S., Y.~Masatlioglu, D.~Nakajima, and E.~Y. Ozbay (2017).
\newblock When more is less: Limited consideration.
\newblock {\em Journal of Economic Theory\/}~{\em 170}, 70--85.

\bibitem[\protect\citeauthoryear{Manzini and Mariotti}{Manzini and
  Mariotti}{2007}]{manzini2007sequentially}
Manzini, P. and M.~Mariotti (2007).
\newblock Sequentially rationalizable choice.
\newblock {\em American Economic Review\/}~{\em 97\/}(5), 1824--1839.

\bibitem[\protect\citeauthoryear{Manzini and Mariotti}{Manzini and
  Mariotti}{2012}]{manzini2012categorize}
Manzini, P. and M.~Mariotti (2012).
\newblock Categorize then choose: Boundedly rational choice and welfare.
\newblock {\em Journal of the European Economic Association\/}~{\em 10\/}(5),
  1141--1165.

\bibitem[\protect\citeauthoryear{Manzini and Mariotti}{Manzini and
  Mariotti}{2014}]{manzini2014stochastic}
Manzini, P. and M.~Mariotti (2014).
\newblock Stochastic choice and consideration sets.
\newblock {\em Econometrica\/}~{\em 82\/}(3), 1153--1176.

\bibitem[\protect\citeauthoryear{Manzini, Mariotti, and Ulku}{Manzini
  et~al.}{2021}]{manzini2019sequential}
Manzini, P., M.~Mariotti, and L.~Ulku (2021).
\newblock Sequential approval: A model of "likes", paper downloads and other
  forms of click behaviour.
\newblock Working paper.

\bibitem[\protect\citeauthoryear{Masatlioglu and Nakajima}{Masatlioglu and
  Nakajima}{2013}]{masatlioglu2013choice}
Masatlioglu, Y. and D.~Nakajima (2013).
\newblock Choice by iterative search.
\newblock {\em Theoretical Economics\/}~{\em 8\/}(3), 701--728.

\bibitem[\protect\citeauthoryear{Masatlioglu, Nakajima, and Ozbay}{Masatlioglu
  et~al.}{2012}]{masatlioglu2012revealed}
Masatlioglu, Y., D.~Nakajima, and E.~Y. Ozbay (2012).
\newblock Revealed attention.
\newblock {\em American Economic Review\/}~{\em 102\/}(5), 2183--2205.

\bibitem[\protect\citeauthoryear{Masatlioglu and Ok}{Masatlioglu and
  Ok}{2005}]{masatlioglu2005rational}
Masatlioglu, Y. and E.~A. Ok (2005).
\newblock Rational choice with status quo bias.
\newblock {\em Journal of Economic Theory\/}~{\em 121\/}(1), 1--29.

\bibitem[\protect\citeauthoryear{Masatlioglu and Ok}{Masatlioglu and
  Ok}{2014}]{masatlioglu2014canonical}
Masatlioglu, Y. and E.~A. Ok (2014).
\newblock A canonical model of choice with initial endowments.
\newblock {\em The Review of Economic Studies\/}~{\em 81\/}(2), 851--883.

\bibitem[\protect\citeauthoryear{Munro and Sugden}{Munro and
  Sugden}{2003}]{munro2003theory}
Munro, A. and R.~Sugden (2003).
\newblock On the theory of reference-dependent preferences.
\newblock {\em Journal of Economic Behavior \& Organization\/}~{\em 50\/}(4),
  407--428.

\bibitem[\protect\citeauthoryear{Noor}{Noor}{2011}]{noor2011temptation}
Noor, J. (2011).
\newblock Temptation and revealed preference.
\newblock {\em Econometrica\/}~{\em 79\/}(2), 601--644.

\bibitem[\protect\citeauthoryear{Ortoleva}{Ortoleva}{2010}]{ortoleva2010status}
Ortoleva, P. (2010).
\newblock Status quo bias, multiple priors and uncertainty aversion.
\newblock {\em Games and Economic Behavior\/}~{\em 69\/}(2), 411--424.

\bibitem[\protect\citeauthoryear{Rideout}{Rideout}{2021}]{rideout2021choosing}
Rideout, S. (2021).
\newblock Choosing for the right reasons.
\newblock Working paper.

\bibitem[\protect\citeauthoryear{Roberts and Lattin}{Roberts and
  Lattin}{1991}]{roberts1991development}
Roberts, J.~H. and J.~M. Lattin (1991).
\newblock Development and testing of a model of consideration set composition.
\newblock {\em Journal of Marketing Research\/}~{\em 28\/}(4), 429--440.

\bibitem[\protect\citeauthoryear{Rozen}{Rozen}{2010}]{rozen2010foundations}
Rozen, K. (2010).
\newblock Foundations of intrinsic habit formation.
\newblock {\em Econometrica\/}~{\em 78\/}(4), 1341--1373.

\bibitem[\protect\citeauthoryear{Salant and Rubinstein}{Salant and
  Rubinstein}{2008}]{salant2008frames}
Salant, Y. and A.~Rubinstein (2008).
\newblock {(A, f): Choice with Frames}.
\newblock {\em The Review of Economic Studies\/}~{\em 75\/}(4), 1287--1296.

\bibitem[\protect\citeauthoryear{Segal}{Segal}{1990}]{segal1990two}
Segal, U. (1990).
\newblock Two-stage lotteries without the reduction axiom.
\newblock {\em Econometrica\/}~{\em 58\/}(2), 349--377.

\bibitem[\protect\citeauthoryear{Sugden}{Sugden}{2003}]{sugden2003reference}
Sugden, R. (2003).
\newblock Reference-dependent subjective expected utility.
\newblock {\em Journal of Economic Theory\/}~{\em 111\/}(2), 172--191.

\bibitem[\protect\citeauthoryear{Tserenjigmid}{Tserenjigmid}{2020}]{tserenjigmid2020characterization}
Tserenjigmid, G. (2020).
\newblock On the characterization of linear habit formation.
\newblock {\em Economic Theory\/}~{\em 70\/}(1), 49--93.

\bibitem[\protect\citeauthoryear{Tserenjigmid}{Tserenjigmid}{2021}]{tserenjigmid2021order}
Tserenjigmid, G. (2021).
\newblock The order-dependent luce model.
\newblock {\em Management Science\/}~{\em 67\/}(11), 6915--6933.

\bibitem[\protect\citeauthoryear{Wright and Barbour}{Wright and
  Barbour}{1977}]{wright1977}
Wright, P. and F.~Barbour (1977).
\newblock Phased decision strategies: Sequels to an initial screening.
\newblock In M.~K. Starr and M.~Zeleny (Eds.), {\em Multiple Criteria Decision
  Making}, pp.\  91--109. Amsterdam: North-Holland Publishing Co.

\bibitem[\protect\citeauthoryear{Xu and Zhou}{Xu and
  Zhou}{2007}]{xu2007rationalizability}
Xu, Y. and L.~Zhou (2007).
\newblock Rationalizability of choice functions by game trees.
\newblock {\em Journal of Economic theory\/}~{\em 134\/}(1), 548--556.

\bibitem[\protect\citeauthoryear{Yildiz}{Yildiz}{2016}]{yildiz2016list}
Yildiz, K. (2016).
\newblock List-rationalizable choice.
\newblock {\em Theoretical Economics\/}~{\em 11\/}(2), 587--599.

\bibitem[\protect\citeauthoryear{Zhang}{Zhang}{2023}]{zhang2023procedural}
Zhang, M. (2023).
\newblock Procedural expected utility.
\newblock Working paper.

\end{thebibliography}
\end{singlespace}

\newpage{}

\appendix

\section{\label{appx:Proof}Appendix: Proofs}

\paragraph*{Simplifying notations:}

\begin{inparaenum}[(i)]\item$A$: a choice set. \item$\boldsymbol{A}$:
a sequence of choice sets (of any length). \item$\left[\boldsymbol{A}\right]_{t=k}^{l}$:
the subsequence of $\boldsymbol{A}$ including only elements in positions
$k$ through $l$. \item$A\in\boldsymbol{A}$: a choice set $A$
that is in the sequence of choice sets $\boldsymbol{A}$. \item$ABC$:
the sequence of choice sets that starts with choice set $A$, followed
by choice set $B$, and ends with choice set $C$. \item$x$: an
alternative. \item$Ax$: alternative $x$ is chosen from the choice
set $A$, which is the only choice set in the sequence. \item$A^{\ni y}x$:
alternative $x$ is chosen from the choice set $A$, which is the
only choice set in the sequence, and alternative $y$ is in $A$.
\item$AxByC^{\ni x}z$: the sequence of choice sets $ABC$ from which
$x,y,z$ are chosen respectively, and alternative $x$ is also in
$C$. \item$\boldsymbol{A}By$: a sequence of choice sets $\boldsymbol{A}$,
followed by the choice set $B$ from which alternative $y$ is chosen.
\item$\boldsymbol{A}xzBy$: a sequence of choice sets $\boldsymbol{A}$,
from which alternative $x$ is chosen from some choice set $A\in\boldsymbol{A}$
and alternative $z$ is chosen from some choice set $C\in\boldsymbol{A}$
(in no particular order), followed by choice set $B$ from which alternative
$y$ is chosen.\end{inparaenum}

\paragraph*{Implications of axioms:}

\axmref{1}, \axmref{2}, and \axmref{3} imply the following conditions.
\oappxref{supplemental_proofs} \subsecref{Proofs-of-Conditions}
provides detailed proofs. Once \thmref{AAT} is proven, these are
also implied by an AAT representation.
\begin{condition}
\label{cond:a1_chosen-or-default}If $\boldsymbol{A}Bx$, either $Bx$
or $\boldsymbol{A}x$.
\end{condition}
\begin{condition}
\label{cond:a2_chosen-without-history}If $\boldsymbol{A}x$, then
$Bx$ for some $B$.
\end{condition}
\begin{condition}
\label{cond:a3_AxBx-not-ByAy}If $AxBx$ and $x\ne y$, then not $ByAy$.
\end{condition}
\begin{condition}
\label{cond:a4_default-means-switch}If $Ax$ and $\boldsymbol{B}Ay$
where $x\ne y$, then $AxCyAy$ for some $C.$
\end{condition}
\begin{condition}
\label{cond:a5_necessary_switches}If $\boldsymbol{A}x$ and $\boldsymbol{B}y$
where $x\ne y$, then either $\boldsymbol{C}yE^{\ni y}x$ or $\boldsymbol{D}xF^{\ni x}y$.
\end{condition}
\begin{condition}
\label{cond:a6_no-various-switches}Suppose $x\ne y$. \begin{inparaenum}[(1)]\item
If $\boldsymbol{A}yB^{\ni y}x$, then not $\boldsymbol{C}xD^{\ni x}y$.
\item If $\boldsymbol{A}xyBx$, then not $\boldsymbol{C}xyBy$. \item
If $\boldsymbol{A}xyB^{\ni y}x$, then not $\boldsymbol{A}xyD^{\ni x}y$.\end{inparaenum}
\end{condition}
\begin{condition}
\label{cond:a7_unidirectional-P}If $xPy$, then not $yPx$.
\end{condition}
\begin{condition}
\label{cond:a8_if-c(A)=00003Dc(B)}If $\boldsymbol{c}\left(\boldsymbol{A}\right)=\boldsymbol{c}\left(\boldsymbol{B}\right)$,
then $\tilde{c}\left(\boldsymbol{A}\right)\left(D\right)=\tilde{c}\left(\boldsymbol{B}\right)\left(D\right)$
for all $D$.
\end{condition}

\subsection{Proof of Theorem \ref{thm:AAT}}

Fix $c$. As is common, necessity of axioms (\textbf{if}) is straightforward,
so I will focus on showing sufficiency of axioms (\textbf{only if}).
The plan goes as follows. We start by constructing $\succ$ in stage
1, the true underlying preference that the anticipated utility function
represents. Stage 2 shows that the constructed $\succ$ has the desirable
properties to be represented by a utility function. In stage 3, we
construct $\Gamma$ and show that $\left(\succ,\Gamma\right)$ explains
choices. Note that by \condref{a2_chosen-without-history}, $\hat{X}=\left\{ x\in X:x=\tilde{c}\left(\emptyset\right)\left(A\right)\exists A\in\mathcal{A}\right\} $.
Moreover, $|X\backslash\hat{X}|\leq1$ because if $z\in X\backslash\hat{X}$,
then $\left\{ z,x\right\} x$ for all $x\ne z$, which means $x\in\hat{X}$
for all $x\ne z$.

\paragraph{Stage 1, construction of $\succ$}

Consider any pair $x,y\in X$ such that $x\ne y$. If $x\in\hat{X}$
and $y\in X\backslash\hat{X}$, set $x\succ_{P}y$. If $x,y\in\hat{X}$,
suppose WLOG that $\left\{ x,y\right\} x$.
\begin{enumerate}
\item If there exists $A$ such that $Ay\left\{ x,y\right\} x$, we set
$x\succ_{S}y$.
\item If there exists $A$ such that $Ay\left\{ x,y\right\} y$, we set
$y\succ_{D}x$.
\end{enumerate}
\begin{claim*}
Either $x\succ_{S}y$ or $y\succ_{D}x$ and not both. 
\end{claim*}
\begin{proof}
The existence of $Ay$ is guaranteed by $y\in\hat{X}$ and \condref{a2_chosen-without-history},
so either $x\succ_{S}y$ or $y\succ_{D}x$. Suppose for contradiction
both, so for some $A$ we have $Ay\left\{ x,y\right\} x$ and for
some $B$ we have $By\left\{ x,y\right\} y$, but this violates \condref{a8_if-c(A)=00003Dc(B)}.
\end{proof}

\paragraph{Stage 2, properties of $\succ$}

By the claim, $\succ_{S}\cup\succ_{D}$, a subset of $\hat{X}\times\hat{X}$,
is connected and antisymmetric. By construction, $\succ_{P}$, a subset
of $\hat{X}\times X\backslash\hat{X}$, is connected by clearly antisymmetric.
Hence the relation $\succ:=\succ_{S}\cup\succ_{D}\cup\succ_{P}$,
a subset of $X\times X$, is connected and antisymmetric. 
\begin{claim*}
$\succ$ is transitive.
\end{claim*}
\begin{proof}
Take any $x,y,z\in X$. If one of $x,y,z$ is in $X\backslash\hat{X}$
(at most one due to $|X\backslash\hat{X}|\leq1$), say WLOG $x,y\in\hat{X}$
and $z\in X\backslash\hat{X}$, then $x\succ z$ and $y\succ z$ means
no violation of transitivity is possible. Now suppose $x,y,z\in\hat{X}$.
Suppose for contradiction $x\succ y$, $y\succ z$, and $z\succ x$.
Since $x,y,z\in\hat{X}$, each of these $\succ$'s is either $\succ_{S}$
or $\succ_{D}$, which implies $xPy$, $yPz$, and $zPx$ (definition
of $P$ given in \secref{Axioms} prior to \axmref{3}). Suppose
WLOG $\left\{ x,y,z\right\} x$. Since $y,z\in\hat{X}$, \condref{a2_chosen-without-history}
guarantees the existence of $Ay$ and $Bz$. Either $AyBz$ or $BzAy$
(or both). To see this, suppose not $AyBz$, then by \condref{a1_chosen-or-default}
we have $AyBy$, but \condref{a3_AxBx-not-ByAy} implies not $BzAz$,
then by \condref{a1_chosen-or-default} we have $BzAy$. Suppose WLOG
$AyBz$, consider $AyBz\left\{ x,y,z\right\} \alpha$. If $\alpha=x$,
then $xPz$ (due to a switch). If $\alpha=y$, then $yPx$ (due to
the default of $x$). If $\alpha=z$, then $zPy$ (due to a switch).
So there is bound to be a contradiction of \condref{a7_unidirectional-P}.
\end{proof}

\paragraph{Stage 3, model explains choice}

Since $\succ$ on $X$ is a strict total order and $X$ is countable,
let $u:X\rightarrow\mathbb{R}$ be real-valued function such that
$u\left(x\right)>u\left(y\right)$ if and only if $x\succ y$. Moreover,
construct the attention function $\Gamma:\mathcal{A}\rightarrow\mathcal{A}$
by $\Gamma\left(A\right):=\left\{ \tilde{c}\left(\emptyset\right)\left(A\right)\right\} $,
for all $A\in\mathcal{A}$. We check that this model specification
explains choices. Throughout, we use $c_{\text{model}}$ to label
the choice function \emph{given by the model}, and from it $\tilde{c}_{\text{model}}$
the one-shot choice functions.
\begin{claim*}
\label{claim:first-choice-okay}$\tilde{c}\left(\emptyset\right)\left(A\right)=\arg\max_{x\in\tilde{\Gamma}\left(\emptyset\right)\left(A\right)}u\left(x\right)$.
\end{claim*}
\begin{proof}
Due to $\Gamma\left(A\right)=\left\{ \tilde{c}\left(\emptyset\right)\left(A\right)\right\} $
and $\tilde{\Gamma}\left(\emptyset\right)\left(A\right)=\Gamma\left(A\right)$.
\end{proof}
We now show that $\left(u,\Gamma\right)$ explains the entire $c$.
Take any sequence of choice sets $\left(A_{n}\right)\in\mathcal{A}^{\text{\textbf{N}}}$,
and suppose for contradiction that, for some $i$, 
\[
c\left(\left(A_{n}\right)\right)_{i}\ne\underset{x\in\tilde{\Gamma}\left(\left(A_{1},...,A_{i-1}\right)\right)\left(A_{i}\right)}{\arg\max}u\left(x\right).
\]
Let $\boldsymbol{i}$ be the set of all such $i$'s; they correspond
to the set of all choice sets in $\left(A_{n}\right)$ from which
the actual choice is not the same as the model prediction. Denote
the minimum element of $\boldsymbol{i}$ by $i^{*}:=\min\boldsymbol{i}$,
which is well-defined. The earlier claim implies $i^{*}\ne1$. Consider
$i^{*}\geq2$. For notational convenience, denote the\emph{ choice}
and the\emph{ model prediction} by, respectively,
\[
c^{R}:=c\left(\left(A_{n}\right)\right)_{i^{*}}\text{\,\,\,and\,\,\,}c^{P}:=c_{\text{model}}\left(\left(A_{n}\right)\right)_{i^{*}}=\underset{x\in\tilde{\Gamma}\left(\left(A_{1},...,A_{i^{*}-1}\right)\right)\left(A_{i^{*}}\right)}{\arg\max}u\left(x\right).
\]

\begin{claim*}
$\left\{ c^{P},c^{R}\right\} \subseteq\tilde{\Gamma}\left(\left(A_{1},...,A_{i^{*}-1}\right)\right)\left(A_{i^{*}}\right)$.
\end{claim*}
\begin{proof}
By definition, $c^{P}\in\tilde{\Gamma}\left(\left(A_{1},...,A_{i^{*}-1}\right)\right)\left(A_{i^{*}}\right)$.
Also, \condref{a1_chosen-or-default} implies either $c^{R}=\tilde{c}\left(\emptyset\right)\left(A_{i^{*}}\right)$
or $c^{R}\in\boldsymbol{c}\left(\left(A_{1},...,A_{i^{*}-1}\right)\right)$,
and since $i^{*}$ is the first instance of disagreement, we have
$c^{R}\in\tilde{\Gamma}\left(\left(A_{1},...,A_{i^{*}-1}\right)\right)\left(A_{i^{*}}\right)$.
We continue with $\left\{ c^{P},c^{R}\right\} \subseteq\tilde{\Gamma}\left(\left(A_{1},...,A_{i^{*}-1}\right)\right)\left(A_{i^{*}}\right)$.
\end{proof}
\begin{claim*}
Either $c^{P}\succ_{S}c^{R}$ or $c^{P}\succ_{D}c^{R}$, which implies
$c^{P}Pc^{R}$.
\end{claim*}
\begin{proof}
By definition, $c^{R}\in\hat{X}$. Since $c^{P}\in\tilde{\Gamma}\left(\left(A_{1},...,A_{i^{*}-1}\right)\right)\left(A_{i^{*}}\right)$,
and since $i^{*}$ is the first instance of disagreement, either $c^{P}\in\Gamma\left(A_{i^{*}}\right)$
which by the construction of $\Gamma$ implies $c^{P}=\tilde{c}\left(\emptyset\right)\left(A_{i^{*}}\right)$
or $c^{P}\in\boldsymbol{c}\left(\left(A_{1},...,A_{i^{*}-1}\right)\right)$,
each would imply $c^{P}\in\hat{X}$. Since the model predicts $c^{P}$
to be chosen over $c^{R}$ even though $c^{R}$ was paid attention
to, we have $u\left(c^{P}\right)>u\left(c^{R}\right)$, which imply
$c^{P}\succ_{S}c^{R}$ or $c^{P}\succ_{D}c^{R}$ from the construction
of $u$. Then $c^{P}Pc^{R}$ follows from the definition of $P$.
\end{proof}
\begin{claim*}
$c^{R}Pc^{P}$.
\end{claim*}
\begin{proof}
Since $i^{*}$ is the first instance of disagreement, $c^{P}\in\tilde{\Gamma}\left(\left(A_{1},...,A_{i^{*}-1}\right)\right)\left(A_{i^{*}}\right)$
implies either $c^{P}=\tilde{c}\left(\emptyset\right)\left(A_{i^{*}}\right)$
or $c^{P}\in\boldsymbol{c}\left(\left(A_{1},...,A_{i^{*}-1}\right)\right)$.
The formal and $c^{R}$ chosen from $A_{i^{*}}$ implies $c^{R}Pc^{P}$.
The latter and $c^{R}$ chosen over $c^{P}$ after $c^{P}$ was previously
chosen implies $c^{R}Pc^{P}$.
\end{proof}
But $c^{P}Pc^{R}$ and $c^{R}Pc^{P}$ contradict \condref{a7_unidirectional-P}.
We showed that if model prediction and actual choices mismatch for
the first time in a sequence, that necessarily leads to a contradiction;
this means no mismatches can ever happen.

\subsection{Proof of Theorem \ref{thm:norelax}}

A standard utility representation with parameter $u$ paired with
$\Gamma\left(A\right)=A$ for all $A$ will imply the other two, which
is also illustrated in the main text.

\paragraph{Past Independence implies standard utility representation:}

Applying Past Independence iteratively gives $\tilde{c}\left(h\right)\left(A\right)=\tilde{c}\left(\emptyset\right)\left(A\right)$
for all $h$ and $A$. Suppose $\tilde{c}\left(\emptyset\right)\left(A\right)$
and $\tilde{c}\left(\emptyset\right)\left(B\right)$ violate WARP,
then the sequence $\left(A,B,A,B,...\right)$ will violate \axmref{1},
so $\tilde{c}\left(\emptyset\right)$ satisfies WARP and it is standard
that there exists $u:X\rightarrow\mathbb{R}$ such that $\tilde{c}\left(\emptyset\right)\left(A\right)=\arg\max_{x\in A}u\left(x\right)$.
So $\tilde{c}\left(h\right)\left(A\right)=\arg\max_{x\in A}u\left(x\right)$
for all $h$ and $A$.

\paragraph{Full Stability implies Past Independence:}

Past Independence is equivalent to $\tilde{c}\left(h\right)\left(A\right)=\tilde{c}\left(\emptyset\right)\left(A\right)$
for all $h$ and $A$ (sufficiency of Past Independence uses iterative
argument, necessity of Path Independence is straightforward). Suppose
Past Independence is not satisfied, so there exists $A$ such that
$\tilde{c}\left(\emptyset\right)\left(A\right)=x$ and $\tilde{c}\left(h\right)\left(A\right)=y\ne x$.
Due to AAT, this means $u\left(y\right)>u\left(x\right)$ and $y\in\boldsymbol{c}\left(h\right)$.
The latter invokes \condref{a2_chosen-without-history} to guarantee
existence of $B$ such that $\tilde{c}\left(\emptyset\right)\left(B\right)=y$.
Consider the sequence of choice set $\left(A,B,\left\{ x,y\right\} ,...\right)$.
In AAT, $x$ is chosen from $A$, $y$ is chosen from $B$ (because
$u\left(y\right)>u\left(z\right)$ for all $z\in\Gamma\left(B\right)\cup\left\{ x\right\} $),
and $y$ is chosen from $\left\{ x,y\right\} $ (because $y\in\tilde{\Gamma}\left(\left(A,B\right)\right)\left(\left\{ x,y\right\} \right)$
and $u\left(y\right)>u\left(x\right)$). But $x$ chosen over $y$
(from $A$) and then $y$ chosen over $x$ (from $\left\{ x,y\right\} $)
within a sequence violates Full Stability.

\subsection{Proof of Theorem \ref{thm:compatibility}}

\paragraph*{Only if:}

Take any $f\in\mathbb{C}$. Since $\mathbb{C}$ is WARP-convex, consider
the $\kappa\in\mathbb{C}_{WARP}$ such that every $\kappa$-cousin
of $f$ is in $\mathbb{C}$. Construct the AAT representation $\left(u,\Gamma\right)$
where $u$ represents $\kappa$ in standard utility maximization and
$\Gamma\left(A\right)=\left\{ f\left(A\right)\right\} $. It is clear
that $c_{0}=f$. Now consider any history $h\in\mathcal{A}^{<\boldsymbol{\text{N}}}$,
we show that $\tilde{c}\left(h\right)\in\mathbb{C}$. Consider any
$A\in\mathcal{A}$. By AAT, $\tilde{\Gamma}\left(h\right)\left(A\right)=\Gamma\left(A\right)\cup\left[A\cap\boldsymbol{c}\left(h\right)\right]=\left\{ f\left(A\right)\right\} \cup\left[A\cap\boldsymbol{c}\left(h\right)\right]$,
so $\tilde{c}\left(h\right)\left(A\right)=\arg\max_{x\in\left\{ f\left(A\right)\right\} \cup\left[A\cap\boldsymbol{c}\left(h\right)\right]}u\left(x\right)=\kappa\left(\left\{ f\left(A\right)\right\} \cup\left[A\cap\boldsymbol{c}\left(h\right)\right]\right)$,
where the second equality is due to the fact that $u$ represents
$\kappa$. By setting $T=\boldsymbol{c}\left(h\right)$, we conclude
that $\tilde{c}\left(h\right)$ is a $\kappa$-cousin of $f$, and
therefore $\tilde{c}\left(h\right)\in\mathbb{C}$. 

\paragraph*{If:}

Suppose $\mathbb{C}$ is compatible with AAT but not WARP-convex,
so for some $f\in\mathbb{C}$, there is no $\kappa\in\mathbb{C}_{WARP}$
such that every $\kappa$-cousin of $f$ is in $\mathbb{C}$. Consider
this $f$ and suppose the AAT representation $\left(u,\Gamma\right)$
is such that $c_{0}=f$ and $\tilde{c}\left(h\right)\in\mathbb{C}$
for all $h\in\mathcal{A}^{<\boldsymbol{\text{N}}}$. Since $u$ represents
some $\kappa\in\mathbb{C}_{WARP}$ in standard utility maximization,
consider the $\kappa$-cousin of $f$ that is not in $\mathbb{C}$,
exists because $\mathbb{C}$ is not WARP-convex, and call it $g$,
which is associated with some finite $T\subseteq f\left(\mathcal{A}\right)$.
Now we construct a history $h\in\mathcal{A}^{<\boldsymbol{\text{N}}}$
such that $\boldsymbol{c}\left(h\right)=T$. Consider the alternatives
$x_{1},...,x_{n}$ such that $\left\{ x_{1},...,x_{n}\right\} =T$
and $u\left(x_{i}\right)<u\left(x_{i+1}\right)$ for each $i$. For
each $i$, since $x_{i}\in T\subseteq f\left(\mathcal{A}\right)$,
there exists $A_{i}$ such that $\tilde{c}\left(\emptyset\right)\left(A_{i}\right)=x_{i}$,
which means $x_{i}\in\Gamma\left(A_{i}\right)$ and $x_{i}=\arg\max_{z\in\Gamma\left(A_{i}\right)}u\left(z\right)$.
Consider the history $h=\left(A_{1},...,A_{n}\right)$. Note that
$\tilde{c}\left(\emptyset\right)\left(A_{1}\right)=x_{1}$. Then,
inductively, for each $i=2,...,n$, $\tilde{\Gamma}\left(\left(A_{1},...,A_{i-1}\right)\right)\left(A_{i}\right)=\Gamma\left(A_{i}\right)\cup\left\{ x_{1},...,x_{i-1}\right\} $
and $u\left(x_{k}\right)<u\left(x_{i}\right)$ for all $k<i$, so
$\tilde{c}\left(\left(A_{1},...,A_{i-1}\right)\right)\left(A_{i}\right)=x_{i}$.
So $\boldsymbol{c}\left(h\right)=T$, and therefore $\tilde{c}\left(h\right)\left(A\right)=\arg\max_{z\in\Gamma\left(A\right)\cup\left[A\cap\boldsymbol{c}\left(h\right)\right]}u\left(z\right)=\arg\max_{z\in\left\{ f\left(A\right)\right\} \cup\left[A\cap T\right]}u\left(z\right)=g\left(A\right)$
for all $A\in\mathcal{A}$, so $\tilde{c}\left(h\right)=g$. It is
impossible that $\tilde{c}\left(h\right)\in\mathbb{C}$ but $g\notin\mathbb{C}$.

\subsection{Proof of Theorem \ref{thm:AATF}}

The proof is identical to the proof of \thmref{AAT}, with these differences:\textbf{
(1)} Whenever the proof of \thmref{AAT}, including its preceding
lemmas and up to stage 3, uses a choice set $A\in\mathcal{A}$, switch
it to a choice problem $\left(A,F\right)\in\bar{\mathcal{A}}$. If
there are multiple of them, $\left(A,F\right),\left(A,F'\right)$,
it is WLOG to pick one. This is where the assumption that $\bar{\mathcal{A}}$
includes every choice set $A\in\mathcal{A}$ at least once becomes
relevant. Note that the definition of $\hat{X}$ now takes into account
frames; that is, if there is any choice problem $\left(A,F\right)$
such that when it appears in any sequences the alternative $x$ is
chosen, then $x\in\hat{X}$.\textbf{ (2)} Then at stage 3, where a
complete and transitive $\succ$ on $X$ has already been established,
proceed with building a utility function $u:X\rightarrow\mathbb{R}$
as usual. Now construct the attention function by $\Gamma\left(A,F\right):=\left\{ \tilde{c}\left(\emptyset\right)\left(A,F\right)\right\} $.\textbf{
(3)} The rest of the proof shows that choices coincide with model
prediction. When it says, ``for all choice set $A\in\mathcal{A}$'',
change it to ``for all choice problem $\left(A,F\right)\in\bar{\mathcal{A}}$''.

\newpage{}

~{\huge{}\vskip 10em}{\huge\par}
\begin{center}
{\huge{}}%
\noindent\begin{minipage}[c]{1\columnwidth}%
\begin{center}
\textbf{\Huge{}Supplemental Materials}{\Huge\par}
\par\end{center}
\begin{center}
\textbf{\Huge{}(Online)}{\Huge\par}
\par\end{center}%
\end{minipage}{\huge\par}
\par\end{center}

\newpage{}

\section{\label{oappx:supplemental_proofs}Online Appendix: Omitted Proofs
and Results}

Simplifying notations are given in \appxref{Proof}.

\subsection{Omitted Proofs}

\subsubsection{\label{subsec:Proofs-of-Conditions}Proofs of Conditions}
\begin{lem}
\label{lem:chosen-or-default}If $c$ satisfies \axmref{2}, then
it satisfies \condref{a1_chosen-or-default}.
\end{lem}
\begin{proof}
Take $\boldsymbol{A}Bx$, and suppose not $Bx$. Let $K$ be the length
of $\boldsymbol{A}$. By \axmref{2}, either $\left[\boldsymbol{A}\right]_{t=1}^{K-1}\left[\boldsymbol{A}\right]_{t=K}^{K}x$
or $\left[\boldsymbol{A}\right]_{t=1}^{K-1}Bx$ (or both). If it is
the former, we are done since $\boldsymbol{A}x$. Suppose it is the
latter, then by \axmref{2} again we have either $\left[\boldsymbol{A}\right]_{t=1}^{K-2}\left[\boldsymbol{A}\right]_{t=K-1}^{K-1}x$
or $\left[\boldsymbol{A}\right]_{t=K}^{K-2}Bx$ (or both). Again,
if it is the former, we are done, otherwise we keep moving backward
until we find $q$ such that $1\leq q<K$ and $\left[\boldsymbol{A}\right]_{t=1}^{q}\left[\boldsymbol{A}\right]_{t=q+1}^{q+1}x$.
If this process does not end when $q=1$, then $\left[\boldsymbol{A}\right]_{t=1}^{1}x$
by \axmref{2}, so $\boldsymbol{A}x$.
\end{proof}
\begin{lem}
\label{lem:chosen-without-history}If $c$ satisfies \axmref{2},
then it satisfies \condref{a2_chosen-without-history}.
\end{lem}
\begin{proof}
Say $\boldsymbol{A}x$, and in particular $x$ is chosen from the
$K$-th element, i.e., $\left[\boldsymbol{A}\right]_{t=1}^{K-1}\left[\boldsymbol{A}\right]_{t=K}^{K}x$.
By \condref{a1_chosen-or-default}, either $\left[\boldsymbol{A}\right]_{t=K}^{K}x$
or $\left[\boldsymbol{A}\right]_{t=1}^{K-1}x$. If the former, let
$B=\left[\boldsymbol{A}\right]_{t=K}^{K}$ and we are done. If the
latter, by \condref{a1_chosen-or-default} again, either $\left[\boldsymbol{A}\right]_{t=K-1}^{K-1}x$
or $\left[\boldsymbol{A}\right]_{t=1}^{K-2}x$. If the former, let
$B=\left[\boldsymbol{A}\right]_{t=K-1}^{K-1}$ and we are done. Otherwise
we keep going backward until we find $q$ such that $1\leq q<K$ and
$\left[\boldsymbol{A}\right]_{t=q}^{q}x$. If this process does not
end by $q=2$, then it must be that $\left[\boldsymbol{A}\right]_{t=1}^{1}x$,
so $Bx$ where $B=\left[\boldsymbol{A}\right]_{t=1}^{1}$.
\end{proof}
\begin{lem}
If $c$ satisfies \axmref{3}, then it satisfies \condref{a3_AxBx-not-ByAy}.
\end{lem}
\begin{proof}
Suppose for contradiction $AxBx$ and $ByAy$. So $\left\{ x,y\right\} \subseteq A\cap B$.
By definition, $By$ and $AxBx$ jointly imply $xPy$. Then $Ax$,
$ByAy$, and $xPy$ jointly violate \axmref{3}.
\end{proof}
\begin{lem}
If $c$ satisfies \axmref{2} and \axmref{3}, then it satisfies \condref{a4_default-means-switch}
\end{lem}
\begin{proof}
Suppose $Ax$ and $\boldsymbol{B}Ay$, then $yPx$ by the definition
of $P$. Moreover \condref{a2_chosen-without-history} implies $Cy$
for some $y$. Now consider $AxC\alpha A\beta$. \condref{a1_chosen-or-default}
and $Cy$ imply $\alpha\in\left\{ x,y\right\} $, but $\alpha=x$,
$Cy$, and $yPx$ jointly contradict \axmref{3}, so $\alpha=y$.
\condref{a1_chosen-or-default} and $Ax$ imply $\beta\in\left\{ x,y\right\} $.
If $\beta=x$, then $xPy$, but $Ax$, $\boldsymbol{B}Ay$, and $xPy$
jointly contradict \axmref{3}. So $AxCyAy$.
\end{proof}
\begin{lem}
If $c$ satisfies \axmref{2} and \axmref{3}, then it satisfies \condref{a5_necessary_switches}.
\end{lem}
\begin{proof}
Suppose $\boldsymbol{A}x$ and $\boldsymbol{B}y$, then by \condref{a2_chosen-without-history}
there exist $Ax$ and $By$. Suppose WLOG $\left\{ x,y\right\} x$.
Consider $ByA\alpha$. \condref{a1_chosen-or-default} implies $\alpha\in\left\{ x,y\right\} $.
Suppose $\alpha=x$, then consider $ByAx\left\{ x,y\right\} \beta$.
No matter $\beta$ we are done. Suppose $\alpha=y$, then \condref{a1_chosen-or-default}
and \condref{a3_AxBx-not-ByAy} imply $AxBy$. Now consider $AxBy\left\{ x,y\right\} \beta$.
No matter $\beta$ we are done.
\end{proof}
\begin{lem}
If $c$ satisfies \axmref{1} and \axmref{3}, then it satisfies \condref{a6_no-various-switches}.
\end{lem}
\begin{proof}
(2) and (3) are immediate given (1), which I now show. Suppose for
contradiction $\boldsymbol{A}yB^{\ni y}x$ and $\boldsymbol{C}xD^{\ni x}y$,
and suppose WLOG $\left\{ x,y\right\} x$. By \axmref{1}, $\boldsymbol{C}xD^{\ni x}y\left\{ x,y\right\} y$.
But this along with $\left\{ x,y\right\} x$ and $xPy$ (by the definition
of $P$ due to $\boldsymbol{A}yB^{\ni y}x$) jointly violate \axmref{3}.
\end{proof}
\begin{lem}
If $c$ satisfies \axmref{1}, \axmref{2}, and \axmref{3}, then
it satisfies \condref{a7_unidirectional-P}.
\end{lem}
\begin{proof}
Suppose $xPy$ and $yPx$. By the definition of $P$, either (i) $\boldsymbol{A}yB^{\ni y}x$
and $\boldsymbol{C}xD^{\ni x}y$, (ii) $\boldsymbol{A}yB^{\ni y}x$
and {[}$Cx$ and $\boldsymbol{D}Cy${]}, or (iii) {[}$Ay$ and $\boldsymbol{B}Ax${]}
and {[}$Cx$ and $\boldsymbol{D}Cy${]} (a fourth case is WLOG the
first case and is omitted). Case (i) contradicts \condref{a6_no-various-switches}.
Due to \condref{a4_default-means-switch}, (ii) and (iii) also contradict
\condref{a6_no-various-switches}.
\end{proof}
\begin{lem}
If $c$ satisfies \axmref{1}, \axmref{2}, and \axmref{3}, then
it satisfies \condref{a8_if-c(A)=00003Dc(B)}.
\end{lem}
\begin{proof}
Suppose $Dz$. Let $c_{1}=\tilde{c}\left(\boldsymbol{A}\right)\left(D\right)$
and $c_{2}=\tilde{c}\left(\boldsymbol{B}\right)\left(D\right)$, and
suppose for contradiction $c_{1}\ne c_{2}$. Say $c_{1},c_{2}\in\boldsymbol{c}\left(\boldsymbol{A}\right)$,
then \condref{a6_no-various-switches} (3) is violated. Instead, suppose
WLOG $c_{1}\notin\boldsymbol{c}\left(\boldsymbol{A}\right)$, so $c_{1}=z$
by \condref{a1_chosen-or-default} and $c_{2}\ne z$, which means
$c_{2}Pz$ and, by \condref{a1_chosen-or-default}, $c_{2}\in\boldsymbol{c}\left(\boldsymbol{A}\right)$.
So $\boldsymbol{A}c_{2}D^{\ni c_{2}}z$, which means $zPc_{2}$. But
$c_{2}Pz$ and $zPc_{2}$ jointly contradict \condref{a7_unidirectional-P}.
\end{proof}

\subsubsection{Proof of Proposition \ref{prop:uniqueness}}

Suppose $x,y\in\hat{X}$ and suppose for contradiction $c$ admits
AAT representations with specifications $\left(u_{1},\Gamma_{1}\right)$
and $\left(u_{2},\Gamma_{2}\right)$ but $u_{1}\left(x\right)>u_{1}\left(y\right)$
and $u_{2}\left(x\right)<u_{2}\left(y\right)$. Since $x,y\in\hat{X}$,
\condref{a5_necessary_switches} guarantees either $\boldsymbol{C}yE^{\ni y}x$,
which contradicts $u_{2}\left(y\right)>u_{2}\left(x\right)$, or $\boldsymbol{D}xF^{\ni x}y$,
which contradicts $u_{1}\left(x\right)>u_{1}\left(y\right)$.

\subsubsection{Proof of Proposition \ref{prop:convergence}}

Suppose $\boldsymbol{C}Ax$ and $\boldsymbol{C}By$. Suppose $u\left(x\right)>u\left(y\right)$.
Since $\tilde{\Gamma}\left(\boldsymbol{C}A\right)\left(B\right)=\tilde{\Gamma}\left(\boldsymbol{C}\right)\left(B\right)\cup\left\{ x\right\} $
and $u\left(x\right)>u\left(y\right)>u\left(z\right)$ for all $z\in\tilde{\Gamma}\left(\boldsymbol{C}\right)\left(B\right)\backslash\left\{ x,y\right\} $,
so $\tilde{c}\left(\boldsymbol{C}A\right)\left(B\right)=x$. Since
$\tilde{\Gamma}\left(\boldsymbol{C}B\right)\left(A\right)=\tilde{\Gamma}\left(\boldsymbol{C}\right)\left(A\right)\cup\left\{ y\right\} $
and $u\left(x\right)>u\left(z\right)$ for all $z\in\tilde{\Gamma}\left(\boldsymbol{C}\right)\left(A\right)\backslash\left\{ x\right\} $
and $u\left(x\right)>u\left(y\right)$, so $\tilde{c}\left(\boldsymbol{C}B\right)\left(A\right)=x$.
We showed convergence on $x$ when $u\left(x\right)>u\left(y\right)$.
If $u\left(y\right)>u\left(x\right)$ instead, analogous arguments
yield convergence on $y$.

\subsubsection{Proof of Proposition \ref{prop:switches}}

\paragraph*{(1)}

Suppose $x\mathbb{S}_{\left(A_{n}\right)}y$, which by definition
means $\boldsymbol{A}yB^{\ni y}x$ exists. Due to AAT, $y\in\boldsymbol{c}\left(\boldsymbol{A}\right)\subseteq\tilde{\Gamma}\left(\boldsymbol{A}\right)\left(B\right)$,
and so $\tilde{c}\left(\boldsymbol{A}\right)\left(B\right)=x$ implies
$u\left(x\right)>u\left(y\right)$.

\paragraph{(2)}

Since $x,y\in\hat{X}$, \condref{a5_necessary_switches} guarantees
either $\boldsymbol{C}yE^{\ni y}x$, which means $x\mathbb{S}y$,
or $\boldsymbol{D}xF^{\ni x}y$, which means $y\mathbb{S}x$. Moreover,
$x\mathbb{S}y$ and $u\left(y\right)>u\left(x\right)$ jointly contradict
AAT, so $x\mathbb{S}y$ if and only if $u\left(x\right)>u\left(y\right)$
for all $x,y\in\hat{X}$, hence $\mathbb{S}$ is also asymmetric and
transitive on $\hat{X}$.

\paragraph{(3) }

Suppose AAT representation $\left(u,\Gamma\right)$ that represents
$c$. Since $\hat{X}$ is finite, enumerate the alternatives in $\hat{X}$
by $u\left(\cdot\right)$ so that $\left\{ x_{1},...,x_{n}\right\} =\hat{X}$
and $u\left(x_{i}\right)<u\left(x_{i+1}\right)$ for all $i$. For
each $i$, \condref{a2_chosen-without-history} guarantees existence
of $A_{i}$ such that $c_{0}\left(A_{i}\right)=x_{i}$. Now construct
the sequence that begins with $\left(A_{1},...,A_{n}\right)$, followed
by the finite sequence of all possible binary choice problems $\left(B_{1},...,B_{k}\right)$
(in any order), with arbitrary completion of what happens next (since
$\left(A_{n}\right)$ must be an infinite sequence). Note that $c_{0}\left(A_{1}\right)=x_{1}$.
For $j\in\left\{ 2,...,n\right\} $, if $\tilde{\Gamma}\left(\left(A_{1},...,A_{j-1}\right)\right)\left(A_{j}\right)\subseteq\Gamma\left(A_{j}\right)\cup\left\{ x_{1},...,x_{j-1}\right\} $,
then $\tilde{c}\left(\left(A_{1},...,A_{j-1}\right)\right)\left(A_{j}\right)=\arg\max_{x\in\tilde{\Gamma}\left(\left(A_{1},...,A_{j-1}\right)\right)\left(A_{j}\right)}u\left(x\right)=x_{j}$.
By induction, this gives $\boldsymbol{c}\left(\left(A_{1},...,A_{n}\right)\right)=\hat{X}$.
Then, in the $\left(B_{1},...,B_{k}\right)$ phase, either $x\mathbb{S}_{\left(A_{n}\right)}y$
or $y\mathbb{S}_{\left(A_{n}\right)}x$ for all $x,y\in\hat{X}$ and
$x\ne y$. Since $x\mathbb{S}_{\left(A_{n}\right)}y$ if and only
if $u\left(x\right)>u\left(y\right)$ for all $x,y\in\hat{X}$, $\mathbb{S}_{\left(A_{n}\right)}$
is also asymmetric and transitive on $\hat{X}$.

\subsubsection{Proof of Proposition \ref{prop:GammaPlus}}
\begin{cor}
\label{cor:subsetgamma}If $c$ admits an AAT representation $\left(\Gamma,u\right)$
and $\Gamma^{*}\subseteq\Gamma$ such that $c_{0}\left(A\right)\in\Gamma^{*}\left(A\right)$
for all $A$, then $c$ also admits an AAT representation $\left(\Gamma^{*},u\right)$.
\end{cor}
\begin{proof}
Denote the resulting choices from $\left(\Gamma^{*},u\right)$ by
$c^{*}$, $\tilde{c}^{*}$, and $\boldsymbol{c}^{*}$. Consider $h=\emptyset$.
For $A$ and $y\in\Gamma^{*}\left(A\right)$, since $y\in\Gamma\left(A\right)$,
so $u\left(\tilde{c}\left(\emptyset\right)\left(A\right)\right)>u\left(y\right)$
if $\tilde{c}\left(\emptyset\right)\left(A\right)\ne y$. Then $\tilde{c}\left(\emptyset\right)\left(A\right)\in\Gamma^{*}\left(A\right)$
implies $\tilde{c}^{*}\left(\emptyset\right)\left(A\right)=\tilde{c}\left(\emptyset\right)\left(A\right)$.
Then, analogous arguments apply to $h\ne\emptyset$ inductively. For
any history $h$, suppose $\boldsymbol{c}\left(h\right)=\boldsymbol{c}^{*}\left(h\right)$
(by induction). For any $A$,
\[
\tilde{\Gamma}^{*}\left(h\right)\left(A\right)=\Gamma^{*}\left(A\right)\cup\left[A\cap\boldsymbol{c}^{*}\left(h\right)\right]\subseteq\Gamma\left(A\right)\cup\left[A\cap\boldsymbol{c}\left(h\right)\right]=\tilde{\Gamma}\left(h\right)\left(A\right).
\]
So for any $y\in\tilde{\Gamma}^{*}\left(h\right)\left(A\right)$,
since $y\in\tilde{\Gamma}\left(h\right)\left(A\right)$, so $u\left(\tilde{c}\left(h\right)\left(A\right)\right)>u\left(y\right)$
if $\tilde{c}\left(h\right)\left(A\right)\ne y$. By AAT, either $\tilde{c}\left(h\right)\left(A\right)=\tilde{c}\left(\emptyset\right)\left(A\right)\in\Gamma^{*}\left(A\right)$
or $\tilde{c}\left(h\right)\left(A\right)\in\boldsymbol{c}\left(h\right)\cap A=\boldsymbol{c}^{*}\left(h\right)\cap A$,
so $\tilde{c}\left(h\right)\left(A\right)\in\tilde{\Gamma}^{*}\left(h\right)\left(A\right)$,
so $\tilde{c}^{*}\left(h\right)\left(A\right)=\tilde{c}\left(h\right)\left(A\right)$.
\end{proof}

\paragraph{Only if:}

Fix $c$. Suppose $c$ admits an AAT representation $\left(\Gamma,u\right)$.
Fix any $A\in\mathcal{A}$. Suppose for contradiction $y\in\Gamma\left(A\right)\backslash\Gamma^{+}\left(A\right)$,
so $y\mathbb{S}c_{0}\left(A\right)$ by definition of $\Gamma^{+}$,
then by \propref{switches} (1) we have $u\left(y\right)>u\left(c_{0}\left(A\right)\right)$.
In this case, $\left(\Gamma,u\right)$ gives $\arg\max_{x\in\tilde{\Gamma}\left(\emptyset\right)\left(A\right)}u\left(x\right)\ne c_{0}\left(A\right)$
a contradiction that $\left(\Gamma,u\right)$ represents $c$. So
$\Gamma\left(A\right)\backslash\Gamma^{+}\left(A\right)=\emptyset$,
or $\Gamma\left(A\right)\subseteq\Gamma^{+}\left(A\right)$. It is
straightforward that $c_{0}\left(A\right)\in\Gamma\left(A\right)$,
otherwise $\arg\max_{x\in\tilde{\Gamma}\left(\emptyset\right)\left(A\right)}u\left(x\right)\ne c_{0}\left(A\right)$,
a contradiction that $\left(\Gamma,u\right)$ represents $c$. 

\paragraph{If:}

Fix $c$. Suppose $c$ admits an AAT representation $\left(\Gamma,u\right)$.
If $z\in X\backslash\hat{X}$, suppose WLOG that $u\left(x\right)>u\left(z\right)$
for all $x\in\hat{X}$. By \corref{subsetgamma}, $\left(\Gamma^{*},u\right)$
where $\Gamma^{*}\left(A\right)=\left\{ c_{0}\left(A\right)\right\} $
also represent $c$. Now I show that $\left(\Gamma^{+},u\right)$
also represent $c$, and then the proof is complete by invoking \corref{subsetgamma}.
Suppose for contradiction there exist $\left(A_{n}\right)$ and integer
$k$ such that the model predictions of $\left(\Gamma^{*},u\right)$
and $\left(\Gamma^{+},u\right)$ disagree. Let $i^{*}$ be the integer
that represents the first disagreement in $\left(A_{n}\right)$, let
$h=\left(A_{1},...,A_{i^{*}-1}\right)$, and denote the model predictions
by, respectively,
\[
a:=\underset{x\in\tilde{\Gamma}^{*}\left(h\right)\left(A_{i^{*}}\right)}{\arg\max}u\left(x\right)\text{\,\,\,and\,\,\,}b:=\underset{x\in\tilde{\Gamma}^{+}\left(h\right)\left(A_{i^{*}}\right)}{\arg\max}u\left(x\right).
\]
Since this is the first disagreement in $\left(A_{n}\right)$, $\boldsymbol{c}^{*}\left(h\right)=\boldsymbol{c}^{+}\left(h\right)$.
Furthermore, since $\Gamma^{*}\left(A_{i^{*}}\right)=\left\{ c_{0}\left(A_{i^{*}}\right)\right\} \subseteq\Gamma^{+}\left(A_{i^{*}}\right)$
by construction, $\tilde{\Gamma}^{*}\left(h\right)\left(A_{i^{*}}\right)\subseteq\tilde{\Gamma}^{+}\left(h\right)\left(A_{i^{*}}\right)$.
So the disagreement is caused by $b\in\Gamma^{+}\left(A_{i^{*}}\right)\backslash\left\{ c_{0}\left(A_{i^{*}}\right)\right\} $.
Because $u\left(b\right)>u\left(a\right)$, $b\in\hat{X}$. So by
the definition of $\Gamma^{+}$ we have $c_{0}\left(A_{i^{*}}\right)\mathbb{S}b$.
By \propref{switches} (1) this implies $u\left(c_{0}\left(A_{i^{*}}\right)\right)>u\left(b\right)$,
so it is impossible for $\left(\Gamma^{+},u\right)$ to predict $b$.

\subsubsection{Proof of Proposition \ref{prop:LocalCLA}}

Given $\left(u,\Gamma\right)$ where $\Gamma$ is an attention filter.
Consider any $h$. If $x\notin\tilde{\Gamma}\left(h\right)\left(B\right)$,
then $x\notin\Gamma\left(B\right)$ and $x\notin\boldsymbol{c}\left(h\right)$.
Note that $x\notin\Gamma\left(B\right)$ also implies $\Gamma\left(B\backslash\left\{ x\right\} \right)=\Gamma\left(B\right)$
since $\Gamma$ is an attention filter. So $\tilde{\Gamma}\left(h\right)\left(B\backslash\left\{ x\right\} \right)$$=\Gamma\left(B\backslash\left\{ x\right\} \right)\cup\left[\left(B\backslash\left\{ x\right\} \right)\cap\boldsymbol{c}\left(h\right)\right]$$=\Gamma\left(B\right)\cup\left[B\cap\boldsymbol{c}\left(h\right)\right]$$=\tilde{\Gamma}\left(h\right)\left(B\right)$.
We established (1), and (2) is implied by definition.

\subsubsection{Proof of Proposition \ref{prop:CLA}}

\paragraph*{If:}

Since $c$ admits an AAT representation, it satisfies \axmref{1},
\axmref{2}, and \axmref{3} (\thmref{AAT}). Suppose $\Gamma$ is
an attention filter. If $\tilde{c}\left(\emptyset\right)\left(T\right)=x$
and $\tilde{c}\left(\emptyset\right)\left(T\backslash\left\{ y\right\} \right)\ne x$
where $x\ne y$, it must be that $\Gamma\left(T\right)\ne\Gamma\left(T\backslash\left\{ y\right\} \right)$,
so $y\in\Gamma\left(T\right)$ since $\Gamma$ is an attention filter.
Then, since $\tilde{c}\left(\emptyset\right)\left(T\right)=x$, $u\left(x\right)>u\left(y\right)$.
Since $y\mathbb{S}x$ would have implied $u\left(y\right)>u\left(x\right)$
due to \propref{switches} (1), not $y\mathbb{S}x$. Hence \axmref{CLA}.

\paragraph*{Only if:}

Since $c$ satisfies \axmref{1}, \axmref{2} and \axmref{3}, by
\thmref{AAT} it admits an AAT representation (\thmref{AAT}). Consider
another AAT representation where the attention function is $\Gamma^{+}$,
guaranteed by \propref{GammaPlus}. We show that $\Gamma^{+}$ is
an attention filter. Suppose for contradiction there exists $A$ and
$y$ such that $y\in A\backslash\Gamma^{+}\left(A\right)$ and $\Gamma^{+}\left(A\right)\ne\Gamma^{+}\left(A\backslash\left\{ y\right\} \right)$.
By the definition of $\Gamma^{+}$, $y\notin\Gamma^{+}\left(A\right)$
implies $y\in\hat{X}$ and $\neg c_{0}\left(A\right)\mathbb{S}y$,
and \propref{switches} (2) implies $y\mathbb{S}c_{0}\left(A\right)$.
Next we argue that $c_{0}\left(A\right)\ne c_{0}\left(A\backslash\left\{ y\right\} \right)$.
Suppose for contradiction $c_{0}\left(A\right)=c_{0}\left(A\backslash\left\{ y\right\} \right)$,
then $y\mathbb{S}c_{0}\left(A\right)$ and the definition of $\Gamma^{+}$
would imply $\Gamma^{+}\left(A\right)=\Gamma^{+}\left(A\backslash\left\{ y\right\} \right)$,
a contradiction. So $y\mathbb{S}c_{0}\left(A\right)$ and $c_{0}\left(A\right)\ne c_{0}\left(A\backslash\left\{ y\right\} \right)$,
but they jointly contradict \axmref{CLA}.

\subsubsection{Proof of Proposition \ref{prop:LocalRSM}}

Given $\left(u,\Gamma\right)$ where $\Gamma$ is a shortlist, let
$S$ be the corresponding rationale. Consider any $h$. Define a new
rationale $S^{*}\subseteq X\times X$ by $xS^{*}y$ if $xSy$ and
$y\notin\boldsymbol{c}\left(h\right)$. Then let $\tilde{\Gamma}^{*}$
be the shortlist given by $S^{*}$. Finally we check that $\tilde{\Gamma}^{*}=\tilde{\Gamma}\left(h\right)$.
Fix any $A\in\mathcal{A}.$ If $y\in\tilde{\Gamma}\left(h\right)\left(A\right)$,
then either $y\in\Gamma\left(A\right)$ (so $\neg xSy$ for all $x\in X$)
or $y\in\boldsymbol{c}\left(h\right)$ (so $\neg xS^{*}y$ for all
$x\in X$), so $y\in\tilde{\Gamma}^{*}\left(A\right)$. If $y\in\tilde{\Gamma}^{*}\left(A\right)$,
then $\neg xS^{*}y$ for all $x\in A$, which by definition of $S^{*}$
can be due to (i) $\neg xSy$ for all $x\in A$, which means $y\in\Gamma\left(A\right)$,
or (ii) $y\in\boldsymbol{c}\left(h\right)$. In either case, $y\in\tilde{\Gamma}\left(h\right)\left(A\right)$.
We established (1), and (2) is implied by definition.

\subsubsection{Proof of Proposition \ref{prop:LocalCTC}}

Given $\left(u,\Gamma\right)$ where $\Gamma$ is a coarse-max, let
$S$ be the corresponding rationale. Consider any $h$. Define a new
rationale $S^{*}\subseteq\left[2^{X}\backslash\left\{ \emptyset\right\} \right]\times\left[2^{X}\backslash\left\{ \emptyset\right\} \right]$
with the following rules. If $R'SR''$, then let $R'S^{*}R''$ if
$R''\cap\boldsymbol{c}\left(h\right)=\emptyset$ and let $R'S^{*}\left[R''\backslash\boldsymbol{c}\left(h\right)\right]$
if $R''\cap\boldsymbol{c}\left(h\right)\ne\emptyset$. The operation
removes $\boldsymbol{c}\left(h\right)$ from categories that would
be dominated, while forming new categories $R''\backslash\boldsymbol{c}\left(h\right)$
that are dominated. Note that as a result, if $R'S^{*}R''$, then
$R''\cap\boldsymbol{c}\left(h\right)=\emptyset$. Then let $\tilde{\Gamma}^{*}$
be the coarse-max defined by $S^{*}$. Finally we check that $\tilde{\Gamma}^{*}=\tilde{\Gamma}\left(h\right)$.
Fix any $A\in\mathcal{A}.$ If $y\in\tilde{\Gamma}\left(h\right)\left(A\right)$,
then either $y\in\Gamma\left(A\right)$ (so $\neg R'SR''$ for all
$R',R''\subseteq A$ and $y\in R''$) or $y\in\boldsymbol{c}\left(h\right)$
(so $\neg R'S^{*}R''$ for all $R''\ni y$), so $y\in\tilde{\Gamma}^{*}\left(A\right)$.
If $y\in\tilde{\Gamma}^{*}\left(A\right)$, then $\neg R'S^{*}R''$
for all $R',R''\subseteq A$ and $y\in R''$, which by definition
of $S^{*}$ can be due to (i) $\neg R'SR''$ for all $R',R''\subseteq A$
and $y\in R''$, which means $y\in\Gamma\left(A\right)$, or (ii)
$y\in\boldsymbol{c}\left(h\right)$. In either case, $y\in\tilde{\Gamma}\left(h\right)\left(A\right)$.
We established (1), and (2) is implied by definition.

\subsubsection{Proof of Proposition \ref{prop:frame-list}}

\paragraph{If: }

Suppose $yFc_{0}\left(\bar{A}\right)$ for some $\bar{A}=\left(A,F\right)\in\bar{\mathcal{A}}$
and, for contradiction, $y\mathbb{S}c_{0}\left(\bar{A}\right)$. So
$u\left(y\right)>u\left(c_{0}\left(\bar{A}\right)\right)$ by an adaption
of \propref{switches} (1). But 
\begin{equation}
c_{0}\left(\bar{A}\right)\in\Gamma\left(\bar{A}\right)\text{ and }yFc_{0}\left(\bar{A}\right)\label{eq:list}
\end{equation}
imply $y\in\Gamma\left(\bar{A}\right)$, so the choice without history
from $\bar{A}$ cannot be $c_{0}\left(\bar{A}\right)$ since $y$
brings greater utility and is considered, a contradiction.

\paragraph{Only if: }

The existence of an AATF representation $\left(u^{*},\Gamma^{*}\right)$
is given by \thmref{AATF}. If $\hat{X}=X$ (there is no never chosen
alternatives), let $u:=u^{*}$. If $X\backslash\hat{X}=\left\{ z\right\} $
(there is at most one never chosen alternatives because all binary
choice sets are in $\mathcal{A}$), let $u:=u^{*}$ and then modify
it by setting $u\left(z\right):=\min u\left(a\right)-1$. Given $\bar{A}=\left(A,F\right)\in\bar{\mathcal{A}}$,
let 
\begin{equation}
\Gamma\left(A,F\right):=\left\{ c_{0}\left(\bar{A}\right)\right\} \cup\left\{ y\in A:yFc_{0}\left(\bar{A}\right)\right\} .\label{eq:list2}
\end{equation}
For each $y\ne c_{0}\left(\bar{A}\right)$ included into consideration,
there are two possibilities. If $y\notin\hat{X}$, then $u\left(y\right)<u\left(z\right)$
for all $z\in X$, so including $y$ into consideration when $c_{0}\left(\bar{A}\right)$
is also considered does not affect choice. If $y\in\hat{X}$, an adaption
of \propref{switches} (1) guarantees that either $y\mathbb{S}c_{0}\left(\bar{A}\right)$
or $c_{0}\left(\bar{A}\right)\mathbb{S}y$, but \axmref{frame-list}
rules out the former, so $c_{0}\left(\bar{A}\right)\mathbb{S}y$,
which implies $u\left(c_{0}\left(\bar{A}\right)\right)>u\left(y\right)$,
hence including $y$ into consideration when $c_{0}\left(\bar{A}\right)$
is also considered does not affect choice. By construction, $\Gamma$
satisfies the desired properties.

\subsubsection{Proof of Proposition \ref{prop:frame-rec}}

\textbf{If:} The same as the proof of \propref{frame-list} except
that \eqref{list} is replaced by $y\in F\left(A\right)$.\textbf{
Only if:} The same as the proof of \propref{frame-list} except that
\eqref{list2} is replaced by $\Gamma\left(A,F\right):=\left\{ c_{0}\left(\bar{A}\right)\right\} \cup\left\{ y\in F\left(A\right)\right\} $
and \axmref{frame-list} is replaced by \axmref{frame-rec}.

\subsection{Omitted Examples and Results}
\begin{example}
\label{exa:population}This example discusses the empirical test
of the convergence property introduced in \subsecref{Identification}.
Consider a population of DMs, each with a (deterministic) AAT representation.
For choice sets $\left\{ \left\{ x,y,z\right\} ,\left\{ x,y\right\} \right\} $,
there are $189$ sets of relevant $\left(u,\Gamma\right)$ parameters
(due to $9$ variations in preferences and $21$ variations of preferences).
Instead of brute-forcing it, let's categorize the DMs using what they
would choose in the first period, which creates a partition of these
$189$ sets using observable differences. Let $P_{ij}$ be the fraction
that would have chosen $i$ from $\left\{ x,y,z\right\} $ and $j$
from $\left\{ x,y\right\} $ when they encounter these choice sets
in the first period; $\sum_{i\in\left\{ x,y,z\right\} ,j\in\left\{ x,y\right\} }P_{ij}=1$.
Within the $P_{xy}$ fraction, let $\lambda_{xy}$ be the fraction
that prefers $x$ to $y$ and $\left(1-\lambda_{xy}\right)$ the fraction
that prefers $y$ to $x$; preference for $z$ is irrelevant since
DMs in this group will never choose $z$ in the problems we consider
(see \propref{convergence}). Define $\lambda_{yx}$ similarly. For
the $P_{zx}$ fraction, let $\lambda_{zx}$ be the fraction that prefers
$x$ to $z$ and $\left(1-\lambda_{zx}\right)$ the fraction that
prefers $z$ to $x$. For the $P_{zy}$ fraction, let $\lambda_{zy}$
be the fraction that prefers $z$ to $y$ and $\left(1-\lambda_{zy}\right)$
the fraction that prefers $y$ to $z$.

Now consider a random assignment of the population of DMs into two
groups (treatments). In group $A$, DMs first choose from $\left\{ x,y,z\right\} $
(first period) and then choose from $\left\{ x,y\right\} $ (second
period). In group $B$, $\left\{ x,y\right\} $ first and $\left\{ x,y,z\right\} $
second. Suppose for now $P_{zx}=P_{zy}=0$, for instance when $z$
is a dominated alternative / decoy. Note that if every DM has full
attention, then $P_{xy}=P_{yx}=0$, but we do not assume this.

In group $A$, for first period's choice (from $\left\{ x,y,z\right\} $),
$P_{xx}+P_{xy}$ fraction chooses $x$ and $P_{yy}+P_{yx}$ fraction
chooses $y$, with relative fraction of $x$ to $y$ being $R_{A}^{t=1}=\frac{P_{xx}+P_{xy}}{P_{yy}+P_{yx}}$.
Similarly, in group $B$, for first period's choice (from $\left\{ x,y\right\} $),
$R_{B}^{t=1}=\frac{P_{xx}+P_{yx}}{P_{yy}+P_{xy}}$. If $P_{xy}=P_{yx}=0$,
then $R_{A}^{t=1}=R_{B}^{t=1}$, so $R_{A}^{t=1}\ne R_{B}^{t=1}$
means some DMs have limited attention.

Now we consider second period's choices. In group $A$, for the second
period choice (from ($\left\{ x,y\right\} $), AAT predicts that the
$P_{xx}$ fraction and the $P_{yy}$ fraction will continue to choose
$x$ and $y$ respectively, but the $P_{xy}$ fraction and the $P_{yx}$
fraction now consider both $x$ and $y$ and choose according to their
preferences; for example, for the $P_{xy}$ fraction, $\lambda_{xy}$
fraction will choose $x$ and $\left(1-\lambda_{xy}\right)$ fraction
will choose $y$ (see \propref{convergence}). The same holds for
group $B$. As a result, relative fractions of $x$ to $y$ for group
$A$ and group $B$ both equal
\begin{align}
R_{A}^{t=2} & =R_{B}^{t=2}=\frac{P_{xx}+P_{xy}\lambda_{xy}+P_{yx}\lambda_{yx}}{P_{yy}+P_{xy}\left(1-\lambda_{xy}\right)+P_{yx}\left(1-\lambda_{yx}\right)}.\label{eq:convergence1}
\end{align}
So $R_{A}^{t=2}=R_{B}^{t=2}=R^{t=2}$, i.e., convergence, even if
$P_{xy},P_{yx}>0$. 

If we assume that the $P_{xx}$ fraction and the $P_{yy}$ fraction
genuinely prefer $x$ and $y$ respectively, then $R^{t=2}$ reveals
the overall fraction of DMs who genuinely prefers $x$ relative to
the overall fraction of DMs who genuinely prefers $y$. In general,
it is possible that a DM belonging to the $P_{xx}$ fraction actually
prefers $y$ to $x$ but never considers $y$.

Now suppose $P_{zx},P_{zy}>0$. If we recalculate $R_{A}^{t=2}$ and
$R_{B}^{t=2}$, they become
\begin{equation}
R_{A}^{t=2}=\frac{...+P_{zx}}{...+P_{zy}},\,\,\,\,\,\,R_{B}^{t=2}=\frac{...+P_{zx}\lambda_{zx}}{...+P_{zy}\left(1-\lambda_{zy}\right)},\label{eq:convergence2}
\end{equation}
where $...$ follow \eqref{convergence1}, so in general $R_{A}^{t=2}\ne R_{B}^{t=2}$.
This occurs due to distributional differences in preferences parameters,
and it would occur even with a population of fully standard DMs (i.e.,
$P_{xy}=P_{yx}=0$ and $\lambda_{zx}=\left(1-\lambda_{zy}\right)=0$).
But this richer dataset provides a solution! In group $A$, $P_{zx}\lambda_{zx}$
fraction will exhibit choices $\left(z,x\right)$, and $P_{zy}\left(1-\lambda_{zy}\right)$
fraction will exhibit $\left(z,y\right)$. In group $B$, $P_{zx}\left(1-\lambda_{zx}\right)$
fraction exhibits $\left(x,z\right)$ and $P_{zy}\left(\lambda_{zy}\right)$
fraction exhibits $\left(y,z\right)$. So data allows us to pin down
$P_{zx},P_{zy},\lambda_{zx},\lambda_{zy}$, hence we can check if
\eqref{convergence2} holds!
\end{example}
\begin{example}
\label{exa:impossible_c-independent}Let $X=\left\{ x,y,z\right\} $.
Consider $\left(u_{1},\Gamma_{1}\right)$ where $u_{1}\left(x\right)>u_{1}\left(y\right)$,
$\left\{ A:y\in\Gamma_{1}\left(A\right)\right\} =\left\{ \left\{ x,y,z\right\} \right\} $,
$\Gamma_{1}\left(\left\{ x,y,z\right\} \right)=\left\{ y\right\} $,
$\Gamma_{1}\left(\left\{ x,y\right\} \right)=\left\{ x\right\} $,
then $x,y\in\hat{X}$ and any sequence $\left(A_{n}\right)$ such
that $x\mathbb{S}{}_{\left(A_{n}\right)}y$ must present $\left\{ x,y,z\right\} $
before $\left\{ x,y\right\} $. Now consider $\left(u_{2},\Gamma_{2}\right)$
where $u_{2}\left(x\right)>u_{2}\left(y\right)$, $\left\{ A:y\in\Gamma_{2}\left(A\right)\right\} =\left\{ \left\{ x,y\right\} \right\} $,
$\Gamma_{2}\left(\left\{ x,y\right\} \right)=\left\{ y\right\} $,
$\Gamma_{2}\left(\left\{ x,y,z\right\} \right)=\left\{ x\right\} $,
then $x,y\in\hat{X}$ and any sequence $\left(B_{n}\right)$ that
reveals $x\mathbb{S}_{\left(B_{n}\right)}y$ must present $\left\{ x,y\right\} $
before $\left\{ x,y,z\right\} $.
\end{example}
\begin{example}
\label{exa:notcompatible}The following $\mathbb{C}$ is not compatible
with AAT: Let $X=\left\{ 1,2,3,4\right\} $. Suppose $\mathbb{C}$
consists of all WARP-conforming choice functions (to make this example
non-trivial) and, additionally, a single choice function $f$ where
$f\left(\left\{ 1,2\right\} \right)=1$, $f\left(\left\{ 2,3\right\} \right)=2$,
$f\left(\left\{ 3,4\right\} \right)=3$, $f\left(\left\{ 4,1\right\} \right)=4$,
$f\left(\left\{ 1,2,3\right\} \right)=3$, $f\left(\left\{ 2,3,4\right\} \right)=4$,
$f\left(\left\{ 3,4,1\right\} \right)=1$, and $f\left(\left\{ 4,1,2\right\} \right)=2$
(the other choice sets are irrelevant). No other choice functions
are in $\mathbb{C}$. Suppose for contradiction that $\mathbb{C}$
is compatible with AAT. To accommodate $f$, by the symmetry of $f$,
suppose without loss of generality $u\left(1\right)>u\left(2\right)>u\left(3\right)>u\left(4\right)$.
After history $h=\left(\left\{ 2,3\right\} \right)$, $2\in\tilde{\Gamma}\left(h\right)\left(A\right)$
if $2\in A$, but this does not change the choices $\tilde{f}\left(h\right)\left(\left\{ 4,1,2\right\} \right)=2$
and $\tilde{f}\left(h\right)\left(\left\{ 1,2\right\} \right)=1$,
so $\tilde{f}\left(h\right)$ violates WARP and $\tilde{f}\left(h\right)\ne f$,
hence $\tilde{f}\left(h\right)\notin\mathbb{C}$, a contradiction
that $\mathbb{C}$ is compatible with AAT.
\end{example}
\begin{cor}
\label{cor:singlesequence_c-indpendent}Suppose $X$ is finite. There
exists a sequence of choice sets $\left(A_{n}\right)\in\mathcal{A}^{\boldsymbol{\text{N}}}$
such that for any $c$\textcolor{blue}{{} }that admits an AAT representation,
there exists a subset of alternatives $\bar{X}_{c}\subseteq X$ such
that $\mathbb{S}_{\left(A_{n}\right)}$ on $\bar{X}_{c}$ is a strict
total order and $\left|X\backslash\bar{X}_{c}\right|\leq1$.
\end{cor}
\begin{proof}
We prove by construction. Consider any sequence $\left(A_{n}\right)$
that begins with the finite sequence of all possible binary choice
problems $\left(B_{1},...,B_{k}\right)$ (in any order) and then repeats
itself, with arbitrary completion of what happens next (since $\left(A_{n}\right)$
must be an infinite sequence). Now consider any $c$. Note that all
but at most one alternative would have been chosen in the first iteration
of $\left(B_{1},...,B_{k}\right)$ (if $z$ is not chosen, then all
other alternatives have been chosen, so $z$ is the only alternative
that has not been chosen), denote this set by $\bar{X}_{c}$. Then,
during the repetition of $\left(B_{1},...,B_{k}\right)$, either $x\mathbb{S}_{\left(A_{n}\right)}y$
or $y\mathbb{S}_{\left(A_{n}\right)}x$ for all $x,y\in\bar{X}_{c}$
and $x\ne y$. Moreover, since $c$ admits an AAT representation,
due to \propref{uniqueness} and $\bar{X}_{c}\subseteq\hat{X}$, we
have $x\mathbb{S}_{\left(A_{n}\right)}y$ only if $u\left(x\right)>u\left(y\right)$.
Hence $\mathbb{S}_{\left(A_{n}\right)}$ is also asymmetric and transitive.
\end{proof}
\begin{cor}
\label{cor:notuniqueattention}If $c$ admits AAT representations
$\left(\Gamma,u\right)$ and $\left(\Gamma',u\right)$, then it also
admits AAT representations $\left(\Gamma\cup\Gamma',u\right)$ and
$\left(\Gamma\cap\Gamma',u\right)$.
\end{cor}
\begin{proof}
The case for intersection is shown in \corref{subsetgamma}; note
that an intersection between $\Gamma$ and $\Gamma'$ guarantees that
$\Gamma^{*}:=\Gamma\cap\Gamma'$ satisfies $\Gamma^{*}\subseteq\Gamma$
and $c_{0}\left(A\right)\in\Gamma^{*}\left(A\right)$ for all $A$.
For the case for union, the ``only if'' of \propref{GammaPlus}
guarantees that $c_{0}\left(A\right)\in\Gamma\left(A\right)\subseteq\Gamma^{+}\left(A\right)$
and $c_{0}\left(A\right)\in\Gamma'\left(A\right)\subseteq\Gamma^{+}\left(A\right)$,
and the ``if'' part subsequently guarantees $\left(\Gamma^{*},u^{*}\right)$
where $\Gamma^{*}=\Gamma\cup\Gamma'$ is an AAT representation for
the same behavior $c$. If $u\ne u^{*}$, then by \propref{uniqueness},
$X\backslash\hat{X}=\left\{ z\right\} $ and $u\left(z\right)>u\left(x\right)$
for some $x\in\hat{X}$ (the proof of \propref{GammaPlus} constructs
$u^{*}$ using $u\left(x\right)>u\left(z\right)$). But since $\left(\Gamma,u\right),\left(\Gamma',u\right)$
both represent $c$, if $c_{0}\left(A\right)=x$, then $z\notin\Gamma\left(A\right)\cup\Gamma'\left(A\right)$,
so $\left(\Gamma^{*},u\right)$ also represent $c$. 
\end{proof}
\begin{cor}
\label{cor:counterfactual-warp}Suppose $c$ admits an AAT representation
$\left(u,\Gamma\right)$ and suppose $c_{0}$ is explained by some
strict total order $\left(\succ_{0},X\right)$ (i.e., $c_{0}\left(A\right)\succ_{0}z$
for all $z\in A\backslash\left\{ c_{0}\left(A\right)\right\} $).
There exists history $h$ such that $\tilde{c}\left(h\right)$ violates
counterfactual WARP if and only if there are $x,y,z\in\hat{X}$ such
that $z\succ_{0}x\succ_{0}y$ but $u\left(x\right)>u\left(y\right)>u\left(z\right)$.
\end{cor}
\begin{proof}
\textbf{If: }This is given by \exaref{2}, with $c_{0}\left(\left\{ y,z'\right\} \right)=y$
replaced by $c_{0}\left(A\right)=y$ for some $A$, which is guaranteed
by $y\in\hat{X}$ and \condref{a2_chosen-without-history}.\textbf{
Only if:} Suppose $c_{0}$ satisfies counterfactual WARP and $\tilde{c}\left(h\right)$
fails counterfactual WARP for some $h\in\mathcal{A}^{<\boldsymbol{\text{N}}}$.
So there exist $x,y\in X$ and $A,B\in\mathcal{A}$ such that $\left\{ x,y\right\} \subseteq A\cap B$,
$\tilde{c}\left(h\right)\left(A\right)=x$, and $\tilde{c}\left(h\right)\left(B\right)=y$.
Suppose WLOG $\tilde{c}\left(h\right)$$\left(\left\{ x,y\right\} \right)=x$.
We will separately demonstrate $u\left(x\right)>u\left(y\right)$,
$u\left(y\right)>u\left(c_{0}\left(B\right)\right)$, $c_{0}\left(B\right)\succ_{0}x$,
and $x\succ_{0}y$, which completes the proof. Note that $c_{0}\left(\left\{ x,y\right\} \right)=x$;
otherwise, $c_{0}\left(\left\{ x,y\right\} \right)=y$ implies $y\in\Gamma\left(\left\{ x,y\right\} \right)\subseteq\tilde{\Gamma}\left(h\right)\left(\left\{ x,y\right\} \right)$,
so $\tilde{c}\left(h\right)\left(\left\{ x,y\right\} \right)=x$ means
$u\left(x\right)>u\left(y\right)$ and $x\in\boldsymbol{c}\left(h\right)$,
but then $\text{\ensuremath{\tilde{c}\left(h\right)}}\left(B\right)=y$
is a contradiction. This means $x\succ_{0}y$. Also, since $c_{0}$
satisfies WARP and $x\in B$, $c_{0}\left(B\right)\ne y$, so the
only way we have $\tilde{c}\left(h\right)$$\left(B\right)=y$ is
$u\left(y\right)>u\left(c_{0}\left(B\right)\right)$ and $y\in\boldsymbol{c}\left(h\right)$,
which also means $c_{0}\left(B\right)\ne x$. Because $c_{0}$ satisfies
WARP, $c_{0}\left(B\right)\succ_{0}x$ and $c_{0}\left(B\right)\succ_{0}y$.
Finally, $y\in\boldsymbol{c}\left(h\right)$ and $\tilde{c}\left(h\right)$$\left(\left\{ x,y\right\} \right)=x$
imply $u\left(x\right)>u\left(y\right)$. 
\end{proof}

\end{document}